\PassOptionsToPackage{pdfpagelabels=false}{hyperref}
\documentclass[fleqn,usenatbib]{mnras}

\usepackage[T1]{fontenc}
\usepackage{ae,aecompl}
\usepackage{adjustbox}

\usepackage{multicol}        
\usepackage{multirow}        

\usepackage{epsfig,amstext,floatfig,alltt,fancyhdr,setspace}
\usepackage{graphicx}
\usepackage{lmodern} 
\usepackage{color}
\usepackage{ucs}
\usepackage[utf8x]{inputenc}
\usepackage{url}
\usepackage{mathrsfs}
\usepackage{amsmath} 
\usepackage{amssymb}
\usepackage{appendix}
\usepackage{rjwmath}
\usepackage{varioref}
\usepackage{times}
\usepackage{enumerate}
\usepackage{magazin_abbrev}
\usepackage{upgreek}

\def\ts{\thinspace}
\def\Eqt{Eq.\thinspace}
\def\sect{Sect.\thinspace}
\def\fig{Fig.\thinspace}
\def\tab{Table\thinspace}
\def\App{Appendix.\thinspace}

\def\B{{\cal B}}

\def\P{{\cal P}}

\def\M{{\cal M}}
\def\Cov{\mathbb{C}}

\def\d{{\rm d}}

\def\CM{{\tens{C}}}
\def\OM{{\tens{O}}}
\def\DM{{\tens{D}}}

\def\MM{\tens{M}}

\def\ellp{\ell'}

\def\dD{\delta_{\rm D}}
\def\kb{{\bf k}}
\def\xb{{\bf x}}
\def\deltab{\boldsymbol{\delta}}

\def\thetab{\pmb \vartheta}
\def\tthetab{\pmb \vartheta^\prime}
\def\lb{\pmb \ell}
\def\tb{\pmb \theta}
\def\Lb{\boldsymbol L}
\def\dlg{\delta^{\rm ln}}
\def\dlgh{\hat\delta^{\rm ln}}
\def\tkappa{\hat\kappa}
\def\tgamma{\hat\gamma}
\def\wgamma{\widetilde{\gamma}}
\def\wkappa{\widetilde{\kappa}}
\def\tC{\widetilde{C}}

\def\pirm{{\rm \pi}}
\def\i{{\rm i}}

\title[]{Flat-Sky Pseudo-Cls Analysis for Weak Gravitational Lensing}

\author[M. Asgari et al.]{
Marika Asgari$^{1}$\thanks{E-mail: ma@roe.ac.uk},
Andy Taylor$^{1}$,
Benjamin Joachimi$^{2}$,
Thomas D. Kitching$^{3}$
\\
$^{1}$SUPA, Institute for Astronomy, University of Edinburgh, Royal Observatory, Blackford Hill, Edinburgh, EH9 3HJ, U.K.\\
$^{2}$Department of Physics and Astronomy, University College London, Gower Street, London WC1E 6BT, UK\\
$^{3}$Mullard Space Science Laboratory, University College London, Holmbury St. Mary, Dorking, Surrey, RH5 6NT, UK \\
}

\date{Accepted 2018 May 29. Received 2018 May 29; in original form  2016 December 14}

\pubyear{2018}

\begin{document}

\label{firstpage}
\pagerange{\pageref{firstpage}--\pageref{lastpage}}
\maketitle

\begin{abstract}
We investigate the use of estimators of weak lensing power spectra based on a flat-sky implementation of the  ‘Pseudo-Cl’ (PCl) technique, where the masked shear field is transformed without regard for masked regions of sky. This masking mixes power, and ‘E’-convergence and ‘B’-modes. To study the accuracy of forward-modelling and full-sky power spectrum recovery we  consider both large-area survey geometries, and small-scale masking due to stars and a checkerboard model for field-of-view gaps. The power spectrum for the large-area survey geometry is sparsely-sampled and highly oscillatory, which makes modelling problematic. Instead, we derive an overall calibration for large-area mask bias using simulated fields. The effects of small-area star masks can be accurately corrected for, while the checkerboard mask has oscillatory and spiky behaviour which leads to percent biases. Apodisation of the masked fields leads to increased biases and a loss of information. We find that we can construct an unbiased forward-model of the raw PCls, and recover the full-sky convergence power to within a few percent accuracy for both Gaussian and lognormal-distributed shear fields.  Propagating this through to cosmological parameters using a Fisher-Matrix formalism, we find we can make unbiased estimates of parameters for surveys up to 1,200 deg$^2$ with 30 galaxies per arcmin$^2$, beyond which the percent biases become larger than the statistical accuracy. This implies a flat-sky PCl analysis is accurate for current surveys but a Euclid-like survey will require higher accuracy.
\end{abstract}




\section{Introduction}

Weak gravitational lensing, the distortion of distant galaxy images by the intervening matter, 
provides us with a unique probe of the mass distribution over a large range of scales in the Universe, 
and so is sensitive to the properties of the dark matter and dark energy
(for a comprehensive review of weak lensing see \citealt{BartelmannSchneider01} 
and for a recent review of cosmic shear see \citealt{Kilbinger_review}). 
In addition, it is sensitive to temporal and spatial distortions of spacetime, 
and hence can be used as a probe of gravity. 
One of the primary concerns of cosmology 
is the comparison of dark matter, dark energy and modified gravity models 
and the estimation of their parameters. 
Cosmological model comparison and parameter estimation are usually carried out 
by compressing the data into a form which can be most easily compared with the models.  
If the data are Gaussian distributed, all of the relevant information 
is contained in the 2-point statistics of the data. 
However, nonlinear evolution of the density field generates higher-order correlations, 
which makes 2-point statistics insufficient for capturing the entire information content of the field.
Nevertheless, much of the information is still retained by the 2-point statistics 
and a lot of effort in cosmology has gone into optimally extracting this information.

The 2-point correlations of the weak lensing shear field can 
be estimated directly from the shear signal on the sky, 
or from the harmonic modes from a transformation of the data.
Many cosmic shear studies have focused on the real space shear 2-point correlation functions (2PCFs),
since this statistic is not biased by the sampling of the shear field from the lensed source galaxy images,
for the scales that are available in the survey.
However, points in the 2PCFs are correlated in a way which does depend on the galaxy sampling 
including the mask and survey geometry \citep[see][]{KilbingerSchneider04}.
In addition, the 2PCFs mix linear and nonlinear scales of the shear field. 
These nonlinear scales can be more difficult to model, in particular due to the presence of baryons 
which affects the evolution of structure.
The cosmic shear field can also be decomposed into a convergence (even-parity or E-mode) 
field which is generated by the matter-density field, and a divergence free (odd-parity or B-mode) field which is mainly 
generated by ellipticity noise and systematics for current surveys. 
The 2PCF mix these two modes, however a full separation can be achieved using COSEBIs \citep[]{SEK10}
 which can also be used to restrict the range of scales 
\citep[see][for cosmic shear with COSEBIs]{Kilbinger13,HEH14,AsgariCFHTLenS16}.

In harmonic space the 2-point observable is the shear power spectrum which, 
by definition, must be positive-semidefinite for auto-spectra. 
The spherical harmonic modes on the celestial sphere are uncorrelated due to rotational invariance and homogeneity, 
moreover, the covariance of the linear shear power spectrum on the full-sky 
is diagonal for Gaussian perturbations before masking effects. 
However, for 2-point statistics, the scales with the most cosmological information are in the non-linear 
r\'egime where the shear powers are correlated.
In order to model accurately  these nonlinear regimes we need simulations, 
which are accurate over a finite range in Fourier space, 
and so a harmonic analysis is well-matched to simulated modelling.
The decomposition into $E$ and $B$-modes is also straightforward in harmonic space on the full-sky.

The main drawback of a harmonic analysis of cosmic shear is the effect of the source galaxy sampling. 
As the shear field is sampled by the background source galaxies, 
a shear map is defined by the position of the source galaxies. 
However, for data analysis it is more convenient to bin the shear data for a harmonic analysis 
and define a shear mask on the pixelated field where no source galaxies are detected. 
Stellar images may also contaminate nearby galaxy images so these galaxies must be excluded. 
On the sky, the mask multiplies the shear field, and so in harmonic space it is a convolution. 
The convolution will correlate different scales, 
and will bias the shear power and covariance unless it is accounted for.

This problem is well-known in CMB analysis, where a spherical harmonic analysis is standard. 
One common, and fast, direct measurement of the CMB power spectrum 
is carried out by the Pseudo-Cl (PCl) analysis \citep[][]{Hivon02}. 
This method can rapidly analyse masked  data in spherical harmonic  space 
and has been used to analyse cosmic microwave background (CMB) temperature 
\citep[see][for a full-sky analysis of the CMB temperature]{Planck14}
and polarization data \citep[see][and references therein, for a full-sky analysis
of simulated and QUaD survey data]{Brown05,Brown09}.
While most CMB analysis takes place on the full, curved sky where the spherical harmonic 
decomposition is well-defined, some studies have used a flat-sky analysis 
\citep[see e.g.][for a flat-sky analysis of the QUaD CMB data]{Yasin}. 
The main advantage of  a flat-sky analysis is speed, especially if very small scales are being analysed, 
as Fast Fourier Transforms (FFTs) can be used, but the choice of modes is more poorly defined than a full-sky analysis, 
depending on the size of the patch analysed.

\cite{Hikage11} used a curved PCl method to analyse simulated data for small masks as well as
investigating flat-sky PCl estimation using similar masks.
\cite{Kitching14} used spherical-Bessel transforms of the shear field on 
flat-sky to preform a 3D cosmic shear analysis of CFHTLenS data.
\cite{Kitching12} also used PCls as a tool to estimate the impact of shape
measurement biases, for the GREAT10 Challenge.
Flat-sky PCl analysis has also been applied to data to estimate
the cross-power spectrum of cosmic microwave background and galaxy lensing maps
\citep[see][]{Hand15,Harnois16} and galaxy-galaxy lensing \citep[see][]{Hikage16}.

CMB data is also analysed using Maximum Likelihood estimators
\citep[see] [for example]{Planck14},
which can also be used on shear fields \citep[see][]{Seljak98,HuWhite01}, and first applied
by \cite{Brown03} for the COMBO-17 survey and \cite{Kohlinger16} 
for the CFHTLenS data.
However, these methods may be too slow for current and future surveys where the number of pixels and
the general resolution is high. Other methods exist which estimate the power spectrum indirectly,
using two-point correlation functions \citep[see for example][]{BeckerDES15,Chon04,Szapudi01}.
 Finally, an alternative approach that shares many of the advantages of 
the pseudo-Cl technique, but can more easily treat the masked regions, 
is Bayesian hierarchical modelling \citep{Alsing2016b,Alsing2016a},
although this method is substantially slower than a PCl analysis.

In this paper we study the effects of masking in the PCl approach for weak lensing, 
for both small and large masks, using Gaussian and
lognormal simulated shear fields, on a flat-sky.
We first go through the formalism of PCls in \sect\ref{sec:PClFormalism},
where we explain how the mode mixing can be modelled via a mixing matrix.
In \sect\ref{sec:PCLResults} the resulting pseudo power spectra and the recovered
power spectra are shown and compared with their expected values from theory.
Finally in \sect\ref{sec:PCLError} we propagate the random and mask modelling errors to the
cosmological parameters using a Fisher analysis and check for significant biases.

\section{Formalism}
\label{sec:PClFormalism}

In this section we review the basic formalism and go through 
some of the more important steps taken to calculate the mixing matrix, 
which models the effects of masking on the power spectra. 
The details of these calculations are given in \cite{Yasin}. 
The following formalism is written for a flat-sky approximation.
The formalism here has some differences from the one outlined in \cite{Hikage11}, 
we apply an additional angular averaging to simplify the relations and speed up the calculations.

The convergence can be separated into two real parts $\kappa_{\rm E}$ and $\kappa_{\rm B}$ in real space. 
Weak gravitational lensing can only produce $\kappa_{\rm E}$ up to first order in the Newtonian gravitational potential. 
Hence any $\kappa_{\rm B}$ would come from other effects, including systematic errors and intrinsic galaxy alignments. 
In Fourier space, we can write $\kappa$ in terms of the Fourier transforms of
$\kappa_{\rm E,B}(\thetab)$, 
\begin{equation}
 \tkappa_\pm(\lb)=\tkappa_{\rm E}(\lb) \pm \i \tkappa_{\rm B}(\lb)\;,
\end{equation}
with
\begin{equation}
 \tkappa_{\rm E,B}(\lb)=\int {\rm d}^2 \vartheta \, \kappa_{\rm E,B}(\thetab)
 e^{-\i\lb\cdot\thetab}\;\;,
\end{equation}
where a hat refers to a Fourier-space quantity.
Note that, $\tkappa_{\rm E,B}(\lb)$ are complex quantities.
We can also write $\gamma_\pm$ as,
\begin{equation}
\gamma_\pm(\thetab)= \gamma_1(\thetab)\pm \i \gamma_2(\thetab)\;,
\end{equation}
where  $\gamma_{1,2}$ are the shear components in Cartesian coordinates.
The Fourier transform of $\gamma_\pm$ is,
\begin{equation}
\tgamma_\pm(\lb)= \tgamma_1(\lb)\pm \i \tgamma_2(\lb)\;,
\end{equation}
where $\tgamma_{1,2}(\lb)$ are the Fourier transforms of $\gamma_{1,2}(\thetab)$,
respectively. 
To find the relation between $\tkappa_\pm(\lb)$ 
and $\tgamma_\pm(\lb)$ we note that they are both functions of the lensing potential, 
$\psi_\pm$, via,
\begin{align}
 \label{eq:GnK}
 \gamma_+&=\frac{1}{2}\partial\partial\psi_+\;
 ,~~~~~\gamma_-=\frac{1}{2}\partial^*\partial^*\psi_-\;
 ,~~~~~\kappa_\pm=\frac{1}{2}\partial^2\psi_\pm\;,
\end{align}
where
\begin{equation}
 \psi_\pm=\psi_{\rm E}(\thetab)\pm \i\psi_{\rm B}(\thetab)\;,~~~~~~~~~~~ \psi_-=\psi_+^*\;,
\end{equation}
and 
\begin{equation}
\label{eqPartial}
\partial\equiv\partial_1+\i\partial_2\;,
    ~~~~~~\partial^\ast\equiv\partial_1-\i\partial_2\;
    ~~~~{\rm and}
    ~~~~~~~\partial^2\equiv\partial\partial^\ast\;,
\end{equation}
where $\partial_{1,2}$ are partial derivatives with respect to $\theta_1,\theta_2$. 
Eliminating $\psi$ in \Eqt\eqref{eq:GnK} results in relations between the shear and convergence,
 \begin{equation}
 \label{eq:Laplacian}
   \kappa_+=\partial^\ast\partial^\ast\partial^{-2}\gamma_+
   ~~~~~~~{\rm and}
   ~~~~~~~~~\kappa_-=\partial\partial\partial^{-2}\gamma_-,
 \end{equation}
 where the inverse Laplacian operator is 
 \begin{equation}
  \partial^{-2}\equiv\int\frac{\d^2 \thetab^\prime}{2\pi}\ln |\thetab-\tthetab|\;.
 \end{equation}
In Fourier space the relation between $\kappa$ and $\gamma$ is more straightforward. 
Using the relation between the partial derivatives in real space with their Fourier counterparts,
 \begin{equation}
 \label{eq:PartialF}
  \mathcal{F}(\partial) =\i\hat\ell\:,~~~~~~~~~~\mathcal{F}(\partial^\ast) =\i\hat\ell^\ast\,,
 \end{equation}
where $\mathcal{F}$ refers to a Fourier transform and
 \begin{equation}
  \hat\ell=\ell_x+\i\ell_y\;,~~~~~~~~~~~~~
    \hat\ell^\ast=\ell_x-\i\ell_y\;,
 \end{equation}
we find 
\begin{align}
\label{kappaPm}
 \hat\kappa_+(\lb)&=\hat\ell^\ast \hat\ell^\ast |\hat\ell|^{-2}
 \tgamma_+(\lb)\;,\nonumber \\ 
 \hat\kappa_-(\lb)&=\hat\ell \hat\ell |\hat\ell|^{-2}
 \tgamma_- (\lb)\;.
\end{align}
Simplifying the above equations by substituting for $\hat\ell$ from,
\begin{equation}
 \hat\ell=\ell e^{\i\varphi_\ell}\;,~~~~ {\rm with}~~~~ \ell=|\hat\ell|\;,
\end{equation}
results in
\begin{equation}
\label{kappaGamma}
\tkappa_\pm(\lb)=e^{\mp 2\i\varphi_{\ell}}\tgamma_\pm (\lb)\;,
\end{equation}
where $\varphi_\ell$ is the polar angle of both $\boldsymbol{\ell}$ and $\hat\ell$.
 
\subsection{Masking effects on shear fields}

In any realistic scenario, parts of the images are masked.
Formally we only need to know the position of source galaxies. 
However, analysing a gridded image is significantly faster, 
since Fast Fourier Transforms can be utilized in this case. 
A PCl analysis relies on such gridded fields, 
where any region with no signal resulting from observers' choices 
or faulty and empty pixels produce the mask. 
We can choose to apodise the masked shear field with a smoothing kernel, $S$, 
to avoid sharp mask features, which make the Fourier transform of the mask challenging. 
The masks used in this work, consist of ones and zeros, exclusively. 
However, in practice the detector defect masks are usually smoother,
due to dithering of the observed images. 
If the mask provided by the observer is smooth enough then it will mimic an apodised  
binary mask.

A mask, $W$, has a multiplicative effect on the shear field,
\begin{equation}
\widetilde{\gamma}_\pm(\thetab)=W(\thetab)\gamma_\pm(\thetab)\;,   
\end{equation}
where we assume $W(\thetab)=0$ corresponds to a fully masked region.
Any quantity with a tilde denotes a masked or pseudo quantity from here on.
There are two ways to apodise a mask, one is to convolve the masked shear field, 
$\wgamma(\thetab)$, with $S$,
\begin{equation}
\wgamma_{\pm}^{\rm s}(\thetab)=\int\d^2\thetab' S(\thetab-\thetab') W(\thetab')\gamma_\pm(\thetab')\;.
\end{equation}
The superscript ${\rm s}$ denotes a smoothed quantity.
The other method is to take the mask and smooth its edges with a kernel. 
Note that when this apodisation method is used, 
the mask will maintain its original zeros while 
smoothly transitioning to the unmasked parts, where $W(\thetab)=1$. 
Therefore, using this method enlarges the mask.
The original mask is then replaced by the new apodised mask.
We use this method for apodising the masks in this work.

In Fourier space the shear field is first convolved with the mask 
(we will drop the hat for Fourier counterparts from here on 
for simplicity, e.g. $\tgamma(\lb)\rightarrow\gamma(\lb)$),
\begin{equation}
\label{eq:Pgamma}
 \wgamma_\pm(\lb) = \int \frac{\d^2 \ell^\prime}{(2\pi)^2}
 W(\lb-\lb')\gamma_\pm(\lb')\;,
\end{equation}
 and then multiplied by the smoothing kernel if the first apodisation method is used,
\begin{equation}
\label{eq:ApPgamma}
 \wgamma_{\pm}^{\rm s}(\lb) =S(\lb) \int \frac{\d^2 \ell^\prime}{(2\pi)^2}
 {W}(\lb-\lb')\gamma_\pm(\lb')\;.
\end{equation}
Substituting from \Eqt{\eqref{kappaGamma}} into the above equation 
we can find a relation for the masked $\kappa$,
\begin{align}
&\wkappa_\pm(\lb) =\int \! \! \frac{{\rm d}^2
  \ell'}{(2\pi)^2} {W}(\lb-\lb')\kappa_\pm(\lb')
  e^{\mp 2\i\varphi_{\ell\ell^\prime}}\;\\ \nonumber
  &{\rm and}~~~\wkappa_{\pm}^{\rm s}(\lb)= S(\lb)\wkappa_{\pm}(\lb)\;, 
  ~~~{\rm where}~~~ \varphi_{\ell\ell^\prime}=\varphi_{\ell}-\varphi_{\ell^\prime}.
\end{align}
By adding and subtracting the equations above, 
we can find a relation between the masked and unmasked $\kappa_{\rm E,B}$,
\begin{align}
\label{eq:kappa}
&\wkappa_{\rm E}(\lb)\!=\!\!\int\!\!\frac{\d^2\ell'}{(2\pi)^2}
{W}(\lb-\lb')[\kappa_{\rm E}(\lb')\cos2\varphi_{\ell \ell'}
\!+\kappa_{\rm B}(\lb')\sin2\varphi_{\ell\ell'}]\;,  \nonumber\\
&\wkappa_{\rm E}^{\rm s}(\lb)\!= S(\lb)\wkappa_{\rm E}(\lb)\;,
  \nonumber \\
&\wkappa_{\rm B}(\lb)\!=\!\!\int\!\!\frac{\d^2\ell'}{(2\pi)^2}
{W}(\lb-\lb')[\kappa_{\rm B}(\lb')\cos2\varphi_{\ell \ell'}
\!-\kappa_{\rm E}(\lb')\sin2\varphi_{\ell \ell'}]\;, \nonumber \\
&\wkappa_{\rm B}^{\rm s}(\lb)\!=S(\lb)\wkappa_{\rm B}(\lb)\;,
\end{align}
where $\wkappa_{ \rm E,B}^{\rm s}(\lb)$ are the smoothed and masked E/B-mode 
$\kappa$.
The above relations show that the mask affects the convergence in Fourier space
by mixing some of the E-mode components into the B-modes and vice versa.
Consequently, in order to utilize Fourier space information in cosmic shear analysis, 
the effects of the mask must be modelled. 

\begin{figure*}
  \begin{center}
    \begin{tabular}{c}
      \resizebox{145mm}{!}{\includegraphics{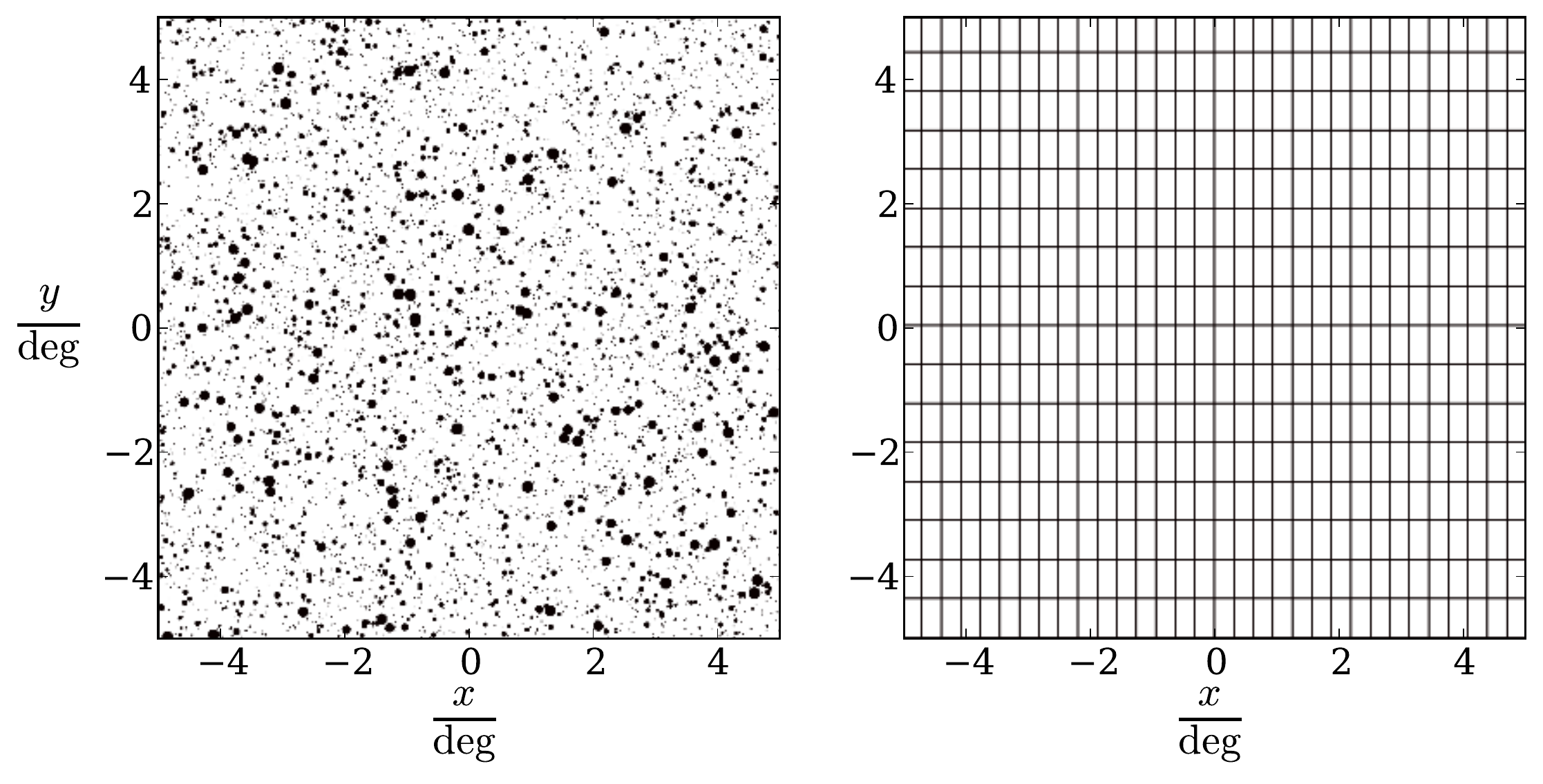}}  
    \end{tabular}
     \caption{\small{Star and checkerboard masks. The star mask contains randomly positioned
		    circles with random areas picked from three ranges, 
		    $2\%$ from $[0.1,0.5]$, $5\%$ from $[1,25]$ 
		    and $3\%$ from $[15,100]$ square arcminutes. 
		    The checkerboard mask mimics a CCD gap pattern. 
		    Three pixels are masked in the gaps. 
		    These two masks are also combined to simulate a more realistic scenario.}}
    \label{figMask}
  \end{center}
\end{figure*}

The masks considered in this work are categorized into two groups: small and large scale masks.
We also combine these masks to make the composite mask. \fig\ref{figMask} shows the star and 
checkerboard masks used throughout this work. These are plausible mask models which resemble masks used for real data 
\citep[see][for example]{Erben13}.
The left panel shows the star mask which contains randomly positioned circles with random areas picked
from three ranges; $2\%$ of the field is covered with stars from $[0.1,0.5]$, 
$5\%$ from $[1,25]$ and $3\%$ from $[15,100]$, square arcminutes.
The checkerboard mask, which represents a CCD chip pattern or any other regular large scale pattern,
contains 3 dark pixels to simulate chip boundaries.
These masks contain only ones and zeros. The masked regions are shown in black.
The masks are zero padded to twice their size in each direction, to minimise artefacts from the assumed periodic boundary conditions.
As can be seen in \fig\ref{figMask} the masks have sharp features which motivates 
smoothing. 

\subsection{The mixing matrix}
\label{sec:Mask}

For a Gaussian isotropic random field all the information is contained in the $C(\ell)$, which only depend on the absolute
value of the wave numbers. However, the mask in general is not isotropic, which means that its power spectrum depends on 
the wave number angle, $\phi_\ell$, as well.
In order to apply the mask to a theory power spectrum, averages over 
its angular dependencies are taken and a mixing matrix is calculated. 
In practice, aside from the mixing matrix the effects of $\ell$-mode binning 
have to be included for a more accurate and comprehensive 
analysis. The exact steps that need to be 
taken in such an analysis are explain in \sect\ref{sec:Binning}. 
For simplicity of the formalism, here we ignore the binning effects. 

The power spectrum is defined as
\begin{equation}
\label{eq:Cl1}
 \left\langle| \kappa_{\rm X}(\lb)\kappa_{\rm Y}^\ast(\lb')|\right\rangle 
 =(2\pi)^2 \dD(\lb -\lb')C(\ell)^{\rm XY}\;,
\end{equation}
where X and Y represent E or B, 
and the angle brackets denote an ensemble average. 
Therefore, the expected average $C(\ell)$ is
\begin{equation}
\label{eq:Cl}
 \left\langle C(\ell)^{\rm XY}\right\rangle=
 \frac{1}{A}\int_0^{2\pi} \frac{\d\varphi_{\ell}}{2\pi}
 \left\langle|\kappa_{\rm X}(\lb)\kappa_{\rm Y}^\ast(\lb)|\right\rangle\;,
\end{equation}
where the integral is a simple angle averaging and over a 
finite area the Dirac delta in \Eqt\eqref{eq:Cl1} is replaced by,
\begin{equation}
\label{eq:DiracDelta}
 \dD(0)\approx\frac{A}{(2\pi)^2}\;,
\end{equation}
where $A$ is the area of the field. The full $\dD$ function is 
recovered when $A\rightarrow\infty$.
In practice because of the existing masks on the images,
we can only measure a pseudo-power spectrum, $ \tC(\ell)$.
In the absence of noise $\tC(\ell)$ is defined in the same 
way as $C(\ell)$ in \Eqt\eqref{eq:Cl}, 
by replacing $\kappa(\lb)$ with the masked convergence, $\wkappa(\lb)$.
However, the cosmological models provide us with the power spectrum, $C(\ell)$.
We can find a relation between $C(\ell)$ and $\tC(\ell)$ by inserting for $\wkappa(\lb)$
from \Eqt\eqref{eq:kappa} into the masked version of \Eqt\eqref{eq:Cl}. 
For example, the E-mode PCl, $\tC^{\rm EE}(\ell)$,
can be written as,
 \begin{align}
 \label{eq:PCL1}
  \left\langle \tC({\ell})^{\rm EE}\right\rangle &=
 \frac{1}{A}\int \frac{\d\varphi_{\ell}}{2\pi} | S(\lb)|^2\\ \nonumber
 &~~~\times\int\!\!\frac{\d^2\ell'}{(2\pi)^2}W(\lb-\lb')
\int\!\!\frac{\d^2\ell''}{(2\pi)^2}W^\ast(\lb-\lb'')\\ \nonumber
&~~~\times\big\langle\big|[\kappa_{\rm E}(\lb')\cos2\varphi_{\ell \ell'}
+\kappa_{\rm B}(\lb')\sin2\varphi_{\ell\ell'}]\\ \nonumber
&~~~~~~~~~\times [\kappa^\ast_{\rm E}(\lb'')\cos2\varphi_{\ell \ell''}
+\kappa^\ast_{\rm B}(\lb'')\sin2\varphi_{\ell\ell''}]\big|\big\rangle\;,
 \end{align}
where $S(\lb)$ can be ignored if the mask is not smoothed or the second apodisation scheme is used.
The mask is not a variable between the realizations (that is assuming that there is no 
correlation between the mask and the underlying shear field), therefore, 
we can take $W$ out of the ensemble averages. 
Moreover, choosing a symmetric smoothing kernel
allows us to take $|S(\lb)|^2$ out of the integral over $\varphi_\ell$.
Using \Eqt\eqref{eq:Cl1} we link the $\tC(\ell)$ to the $C(\ell)$,
\begin{align}
\label{eq:Celldag1}
 \left\langle \tC^{\rm EE}(\ell)\right\rangle &=
 \frac{|S(\ell)|^2}{A}\int \frac{\d\varphi_{\ell}}{2\pi}
 \int\!\!\frac{\d^2\ell'}{(2\pi)^2} |W(\lb-\lb')|^2\\ \nonumber
&\times\big\{C^{\rm EE}(\ell')\cos^2 2\varphi_{\ell \ell'}\\ \nonumber
&~~~+[C^{\rm EB}(\ell')+C^{\rm BE}(\ell')]\sin 2\varphi_{\ell \ell'}\cos 2\varphi_{\ell \ell'}\\ \nonumber
&~~~+C^{\rm BB}(\ell')\sin^2 2\varphi_{\ell \ell'}\big\}\;.
\end{align}
The above equation is written for the E-mode power spectrum although it can be extended to the 
other cases, shown in \App\ref{appMixing}.
While $C(\ell')^{\rm XY}$ only depend on $|\lb'|$,
$W(\lb-\lb')$ and the trigonometric functions in \Eqt\eqref{eq:Celldag1} 
depend on the polar angles $\varphi_{\ellp}$ and $\varphi_{\ell}$. 
Therefore, the angle averaging part of the integrals in \Eqt\eqref{eq:Celldag1} can be taken 
independent of the cosmological model \citep[see][]{Yasin}.
The details of the calculations are given in \App\ref{appMixing}.
The masking effect is hence modelled in the form of a mode mixing matrix, $\M$, 
\begin{equation}
\label{eq:M}
\M({\ell, \ell'}) \equiv\frac{|S(\ell)|^2}{(2\pi)^2 A}\int_{0}^{\pi}\!\!\! 
\d \eta\; W_{\gamma\gamma}\big(L(\ell,\ell',\eta)\big) M_{\upeta}(\eta)\;,            
\end{equation}
where $\eta$ is the angle between $\lb$ and $\lb'$, 
\begin{equation}
 L(\ell,\ell',\eta)\equiv|\lb-\lb'|=\sqrt{\ell^2+\ell'^2-2\ell\ell' \cos(\eta)}\;,
\end{equation}
and $W_{\gamma\gamma}(L)$ is the power spectrum of the mask,
\begin{equation}
\label{eq:Wgg}
 W_{\upgamma\upgamma}(L)\equiv\int_0^{2\pi}\frac{\d\varphi_{L}}{2\pi}|W(\boldsymbol L)|^2,
\end{equation}
and 
\begin{equation}
 M_\upeta(\eta)\equiv\left( \begin{array}{ccc} 1+\cos 4 \eta &  1-\cos 4 \eta & 0\\
                                 1-\cos 4 \eta &  1+ \cos 4 \eta &0 \\
                                       0 & 0  & 2 \cos 4 \eta
                                 \end{array} \right)\;.
\end{equation}
As a result we can write
\begin{equation}
\label{eq:PCL}
 \boldsymbol{\tC}(\ell)= \int_0^\infty\d \ell' \ell'\M(\ell, \ell') \boldsymbol{C}(\ell')\;,
\end{equation}
where
\begin{equation}
  \boldsymbol{C}\equiv\big(C^{\rm EE}(\ell), C^{\rm BB}(\ell), C^{\rm EB}(\ell)\big)^t\;.
\end{equation}
%
In practice we need to change all the integrals in the above equations into discrete finite sums. 
Therefore, we write \Eqt\eqref{eq:PCL} as follows for discrete values of $\ell$,
\begin{equation}
\label{eq:discretePCL}
 \boldsymbol{\tC}(\ell_i)= \sum_j\Delta \ell_j \ell_j\M(\ell_i, \ell_j) \boldsymbol{C}(\ell_j)\;.
\end{equation}
The $\ell_i$ values depend on the size of the real space size of the field and the binning used. 
We explain the details of the binning used in this work in \sect\ref{sec:Binning}. For the above equation
however we use the smallest $\ell$-bin that is allowed, to keep the discrete sum as close as possible to its
continuous form in \Eqt\eqref{eq:PCL}. Note that the integrals in \Eqt\eqref{eq:M} and \Eqt\eqref{eq:Wgg} 
also need to be changed into discrete sums. 
These approximations are the main reason for the biases that we will see in the results sections.
Full-sky PCls also suffers from biases, which are discussed in \cite{Elsner16}, where they offer a proposal to 
resolve them.  A full-sky analysis, however, does not suffer from some of the limitations faced by 
flat-sky analysis are since the limits on the Fourier modes and the binning are well-defined.

We absorb $\Delta \ell_i \ell_i$ in \Eqt\eqref{eq:discretePCL}
in $\M(\ell_i, \ell_j)$ and define a new mixing matrix,
\begin{equation}
 M_{\ell_i \ell_j}=\Delta \ell_j \ell_j \M(\ell_i, \ell_j)\;,
\end{equation}
which satisfies this matrix relation,
\begin{equation}
\label{eq:PCLWithM}
 {\tC}_\ell= M_{\ell \ell'} {C}_{\ell'}\;,
\end{equation}
where, for simplicity, we have dropped the $i$ and $j$ subscripts 
and replaced them with $\ell$ and $\ell'$ instead
and Einstein summation rules apply. 
The ${\tC}_\ell$ are the elements of a vector of $\tC(\ell)$ for discrete values of $\ell$.
Consequently, we can write the inverse of \Eqt\eqref{eq:PCLWithM} 
and recover the power spectrum,
\begin{equation}
\label{eq:CLWithM}
 {C}^{\rm rec}_\ell= (\MM^{-1})_{\ell \ell'} {\tC}_{\ell'}\;,
\end{equation}
provided we can invert $\MM$.

A prominent source of noise in weak lensing analysis is the galaxy shape noise which we model as
a Gaussian random noise with zero mean and $\sigma_\epsilon$ dispersion corresponding to the 
dispersion of the complex galaxy ellipticity \citep[see][]{Hu99}. 
The mask affects the noise in the same way as the shear field. 
We can write the noise as a separate source of power with no $\ell$ dependence, 
\begin{equation}
\label{eq:NoisePower}
 N_\epsilon=\frac{\sigma_\epsilon^2}{2n_{\rm gal}}\;,
\end{equation}
where $n_{\rm gal}=30$ per square arcminutes is the mean number density of galaxies
and $\sigma_\epsilon=0.3$ is the intrinsic dispersion of galaxy ellipticities, 
similar to values expected from a Euclid-like future survey 
\citep[see][]{EuclidRB11}\footnote{www.Euclid-ec.org}.
As a result the measured PCl is,
\begin{equation}
\label{eq:PCLWithMnN}
 \boldsymbol{\tC}= \MM [\boldsymbol{C}+N_\epsilon]\;,
\end{equation}
and the recovered Cl is,
\begin{equation}
\label{eq:CLMat}
 \boldsymbol{C}^{\rm rec}= (\MM^{-1})\boldsymbol{\tC}-N_\epsilon\;.
\end{equation}

\subsection{Mask smoothing: apodisation}
\label{sec:Apodise}

We use three Gaussian smoothing kernels to apodise the masks using the second method 
(smoothing the edges of $W(\thetab)$ before applying it to the shear fields).
Hence in the equations where $S$ appears it should be ignored.
The advantage of this method is that it allows slower variations for the integrands in 
\Eqt\eqref{eq:M} and \Eqt\eqref{eq:Wgg} which could make their discrete approximation more accurate.
As can be seen in \Eqt\eqref{eq:M} in the case of the first apodisation method, the smoothing
kernel only comes into play after the angular averages in \Eqt\eqref{eq:M} and \Eqt\eqref{eq:Wgg} 
have been taken. 

Note that using this method increases the effective masked area,
since the fully masked regions will remain the same, while their edges will have a 
smooth transition from zero to one, which is determined by the size of the kernel.
In general, any smoothing function can be chosen as the kernel.
However, here we use here Gaussian Kernels. 
They are identified using the number of pixels that determines their 
size in real space pixels, $N$, which is an odd number. 
We set all the values outside a box centred at the maximum of the Gaussian
with $N$ pixels on each side to zero and set the dispersion of the Gaussian kernel to $(N-1)/1.5$.
The apodisation is done around the edges of the masks, such that the apodised mask starts from 0 on the edge of the original mask 
and transitions to 1 over roughly N-1 pixels. An odd N is chosen so that the kernel is symmetric around its origin\footnote{The apodisation is done by first zero padding each mask with $(N-1)/2$ pixels and then convolving it with a two dimensional convolution method (filter2 in {\sc matlab}).}. Here we use three sizes for the Gaussian kernels which are listed in \tab\ref{tabAp}.
The main results are shown and compared for the original masks and these three apodisation schemes. 

\begin{table}
\caption{\small{Apodisation case name and the number of pixels, $N$, used in defining the kernel. 
The kernels used in this work are Gaussian functions with $\sigma=(N-1)/1.5$ with a range of support equal to $N$ pixels.
The first row shows the name given to each case. }}
\begin{center}
\begin{tabular}{ | c | c | c | c | c | c | c |  }
  \hline
   & Ap1 & Ap2 & Ap3  \\
  \hline
  $N$  & 5 pixels & 11 pixels &  23  pixels \\
  \hline
\end{tabular}
\end{center}
\label{tabAp}
\end{table}

\subsection{Binning and pixelization effects: theory Cl and PCl}
\label{sec:Binning}

\begin{figure}
  \begin{center}
    \begin{tabular}{c}
      \resizebox{70mm}{!}{\includegraphics{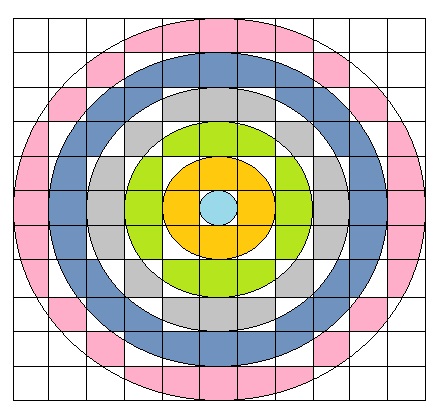}}  
    \end{tabular}
     \caption{\small{A visual presentation of the angle averaging scheme. 
     For the smallest binning scheme each annulus is defined between two adjacent circles. 
     An average is taken over the pixels with centres lying in the annuli, painted in different colours.
     The $\ell$ value that corresponds to the angle averaged quantity is also the average of the $|\lb|$ of the 
     relevant pixels. The binning is done by merging 2 or more of the smallest annuli with each other. } }
    \label{figAngleAve}
  \end{center}
\end{figure}

In practice the 2D Fourier fields (shear and mask) are pixelated,
hence their angle averaged values are not exact and depend on the method used. 
We will ignore the window function of the map pixel shape, as it is 
only important for very high Fourier modes which are not used in this work.
To take the angular average over such a field 
we choose an annulus around the centre of the field, 
identify all the pixels with centres lying inside the annulus
and take their average value. 
The Fourier mode, $\ell$, that corresponds to this estimated value is also calculated 
by averaging over the value of the $|\lb|$ modes that lie in this annulus.
\fig\ref{figAngleAve} demonstrates the angle averaging scheme. 
The edges of the annuli are shown as concentric circles, and the pixels that correspond to an annulus are shown in
different colours. The second annulus for example has 8 pixels with their centres lying inside it which are painted orange.
As we go to larger annuli the number of pixels increases which in turn increases the accuracy of the angle averaging.

To estimate $W_{\gamma\gamma}$ from \Eqt\eqref{eq:Wgg}, we use annuli of width equal to
the smallest Fourier mode, $\Delta\ell_{\rm min}$, available in the field,
\begin{equation}
\label{eq:EllMin}
 \Delta\ell_{\rm min}=\frac{2\pi}{D},
\end{equation}
where $D$ is the side length of the square field after zero padding.
Then the estimated $W_{\gamma\gamma}$ values are fed into \Eqt\eqref{eq:M}. 
Since $W_{\gamma\gamma}$ is only estimated for discrete values, the integral in \Eqt\eqref{eq:M}
needs to be transformed into a sum over these values of $L$. 
This integral is taken over $\eta$, the angle between $\lb$ and $\lb'$,
which is calculated for each $L$, $\ell$ and $\ell'$ from,
\begin{equation}
 \cos \eta= \frac{\ell^2+\ell'^2-L^2}{2\ell\ell'}\;.
\end{equation}
The values of $\ell$ and $\ell'$ depend on the binning scheme used. 
Note that in \Eqt\eqref{eq:M}, the available $\eta$ do not form a regular grid, therefore, 
$\d\eta$ is not constant. 

To find the PCls from theory, first $\MM_{\ell\ell'}$ 
is estimated for the smallest binning that corresponds 
to the field after zero-padding, similar to $W_{\gamma\gamma}$.
Second, the input theory power spectrum, $C^{\rm input}(\ell)$, 
is laid on a grid with the same pixel size and area. Then it is angle averaged,
to find $C^{\rm ave}_\ell$. $C^{\rm input}(\ell)$ and $C^{\rm ave}_\ell$ 
show the largest differences for the smallest $\ell$-modes
as expected, since the number of pixels that lie in the first few annuli 
for the angle averaging is not representative, which biases the results (see \fig\ref{figAngleAve}).
Finally, the theory value of PCls is estimated by applying $\MM_{\ell\ell'}$ to the sum 
of $C^{\rm ave}_\ell$  and $N_\epsilon$. The $\tC(\ell)$ can then be re-binned into broader bins as desired. 
These steps can be summarized in,
\begin{equation}
 \boldsymbol{\tC}^{\rm th}_{\rm b}=\B \{\MM (\boldsymbol{C}^{\rm ave}+N_\epsilon)\}\;,
\end{equation}
where $\boldsymbol{\tC}^{\rm th}_{\rm b}$ is the binned theory PCl 
and $\boldsymbol{C}^{\rm ave}$ is $C^{\rm ave}_\ell$ in vector format,
$\MM$ is the mixing matrix, and $\B$ is the binning matrix defined as follows:

We define $n_\ell$ as the number of smallest bins of size $\Delta\ell_{\rm min}$ 
 that are combined to make the wider bins. 
If the total number of initial $\ell$ bins is $n_{\rm tot}$ then 
there are $n_{\rm bin}=\lfloor n_{\rm tot}/n_\ell \rfloor$
wider bins, where $\lfloor x \rfloor$ denotes the largest integer that is smaller than $x$.
$\B$ is then an $n_{\rm bin}\times n_{\rm tot}$ matrix of this form,
\begin{equation}
\label{eq:Binning}
\B_{b\ell}=\!
\begin{cases} 
  &\!\!\!\!\! n_{\rm p}(\ell)\left(\sum\limits_{(b-1)n_\ell+1}^{b n_\ell} 
  n_{\rm p}(\ell)\right)^{-1}\;\;\; (b-1)n_\ell<\ell\leq b n_\ell \\
  & 0 \;\quad\quad\quad\quad\quad\quad\quad\quad \ell>b n_\ell\quad {\rm or}\quad \ell\leq(b-1)n_\ell
   \end{cases}
\end{equation} 
where $n_{\rm p}(\ell)$ is the number of pixels in each initial $\ell$-bin (see \fig\ref{figAngleAve}).

The ellipticity noise contribution, $\B (\MM N_\epsilon)$, 
can be subtracted from the theory and measured values subsequently.
We forward model the $\tC^{\rm th}_\ell$ so it is the closest to the estimated PCls.

We can use two methods to recover a $C_\ell$, which result in very different values. 
The first method is to apply the inverse of the mixing matrix 
on the $\tC^{\rm est}_\ell$ measured from the fields and then bin the result into wider $\ell$-bins,
\begin{equation}
\label{eq:ClRecI}
{\rm I:\;\; } \boldsymbol{C}^{\rm rec}_{\rm b}=\B(\MM^{-1} \boldsymbol{\tC}^{\rm est})-N_\epsilon\;,
\end{equation}
where $\boldsymbol{C}^{\rm rec}_{\rm b}$ is the binned recovered $C(\ell)$ in vector format.
This recovered power spectrum can be compared with a binned $C^{\rm ave}_\ell$. 
The advantage of this method is that it is less computationally 
intensive as the mixing matrix is only applied once on the
$\tC^{\rm est}_\ell$ and not on the different theory values.

The second method is to recover $C(\ell)$ by applying the inverse of a binned mixing matrix
to a binned estimated $\tC(\ell)$,
\begin{equation}
\label{eq:ClrecII}
{\rm II:\;\; } \boldsymbol{C}^{\rm rec}_{\rm b}=\MM^{-1}_{\rm b} \boldsymbol{\tC}^{\rm est}_{\rm b}\;.
\end{equation}
We can write the predicted theory value for this recovered $C^{\rm rec}_\ell$ as,
\begin{equation}
\label{eq:ClthII}
\boldsymbol{C}^{\rm th}_{\rm b}=\MM^{-1}_{\rm b} \boldsymbol{\tC}^{\rm th}_{\rm b}
               =\MM^{-1}_{\rm b} \B \{\MM (\boldsymbol{C}^{\rm ave}+N_\epsilon)\}\;,
\end{equation}
where the noise contribution $\MM^{-1}_{\rm b} \B \MM N_\epsilon$ can be subtracted from both recovered and theory
values in Eqs.\thinspace\eqref{eq:ClrecII} and \eqref{eq:ClthII}.
To use the second method we would need to apply the mixing matrix on the $C^{\rm ave}_\ell$ value to find 
 $\tC^{\rm th}_\ell$. Therefore, this method is at least as computationally demanding as the forward modelling 
were the theory PCls are compared to their measured values. 
Furthermore, as the binned mixing matrix is more diagonal, the shape of the recovered $C_\ell$ from this method is 
similar to $\boldsymbol{\tC}^{\rm th}_{\rm b}$, instead of the underlying power spectrum.

\subsection{Mask power spectra and mixing matrices}

In the past sections we explained the methods we use to find the power spectrum and mixing 
matrix of a mask, as well as our apodisation scheme. Here we show the mask power spectra, 
$W_{\gamma\gamma}(\ell)$, for all the mask and apodisation combinations and show examples of mixing matrices.

\begin{figure*}
  \begin{center}
    \begin{tabular}{c}
      \resizebox{165mm}{!}{\includegraphics{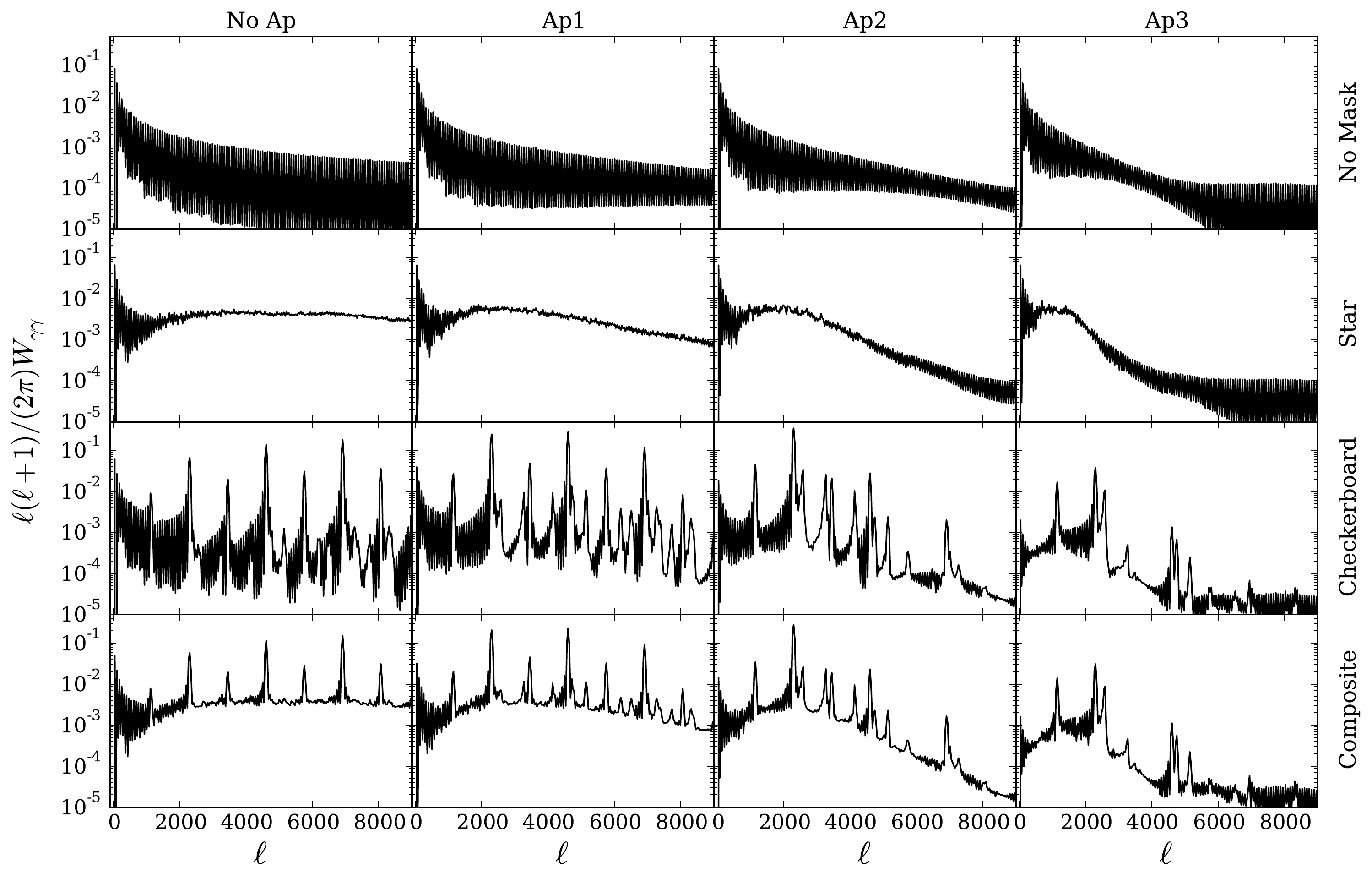}}  
    \end{tabular}
     \caption{\small{Normalized power spectra of all the mask and apodisation combinations 
     for the smallest binning ($\Delta \ell_{\rm min}\approx 18/{\rm rad}$). 
     The fields are all zero-padded, hence even the "No Mask" case has a square shaped mask.
     The composite mask is one that combines the star and checkerboaard patterns. 
     ``No Ap'' means that no apodisation has been applied to the original mask, while 
     ``Ap1'', ``Ap2'' and ``Ap3'' indicate masks which are apodised using Gaussian kernel with 
     increasing support respectively (see \sect\ref{sec:Apodise}). 
     Note that the x-axis is linear while the y-axis is in logarithmic scale.} }
    \label{fig:Wgg}
  \end{center}
\end{figure*}

\fig\ref{fig:Wgg} shows the mask power spectra defined in \Eqt\eqref{eq:Wgg}. 
Each row belongs to a mask configuration, while each column shows the results for different apodisation cases, 
which are indicated at the right hand side of the rows and the top of the columns. 
The fields are zero-padded before the FFT is applied to them, therefore, 
even the "No Mask" case has a large scale square shaped mask, which one could call the survey footprint.
The zero-padding allows for a higher resolution estimate of $W_{\gamma\gamma}$, 
effectively interpolating between the natural $\ell$ values of the original pixelated field.
We zero-pad the masks to double their size on each side. 

$W_{\gamma\gamma}$ is presented for the smallest binning available, 
since it is used in this format to find the mixing matrix
(see \Eqt\ref{eq:M}).
We use the "No Mask" option as the control case. 
By comparing the first row in the figure with the following rows we see that 
the largest scale feature, at small $\ell$ values, is due to the zero-padding. 
The Fourier transform of a perfect square mask is a double sinc function.
Therefore, the power spectrum of such a mask oscillates heavily, which is what we see for 
the ``No Mask'' version. As we add more structures to the mask this oscillatory behaviour is suppressed,
which is what we see in \fig\ref{fig:Wgg}. One of the main differences between the current work and \cite{Hikage11}
is that in contrast to this work, they assumed periodic boundary conditions and only small scale masks to study PCls.
Unlike the star mask which has a relatively featureless power spectrum, 
the checkerboard shows very prominent peaks, due to its regular pattern. 
The star mask shows relatively constant power over a large range of $\ell\gtrsim2000$,
since it consists of randomly positioned stars of different sizes. The star mask is 
basically self-similar for this range of scales.
The features in the power spectrum of the checkerboard mask, 
form a comb corresponding to the harmonics of the regular pattern.
Specifically $\ell~1150$ corresponds to the first harmonic of the checkerboard pattern.

By comparing different columns in \fig\ref{fig:Wgg}, we see that the 
apodisation dampens the tail of the power spectra, 
the scale and strength of which depends on the size of the kernel. 
As a result, we see that ``Ap3'' which is the largest kernel we use has a more dramatic
effect on $W_{\gamma\gamma}$ compared to the smaller kernels. Note that the apodisation here
only smooths the edges of the masked regions while keeping the zeros in the mask intact, hence 
more apodisation results in a larger effective masked area.
The smoothing is not as effective on the checkerboard power spectra, 
which will have important consequences for the estimation of the convergence power spectra, 
the effects of which will become apparent in \sect\ref{sec:FisherResults}. 

\begin{figure*}
  \begin{center}
     \resizebox{145mm}{!}{\includegraphics{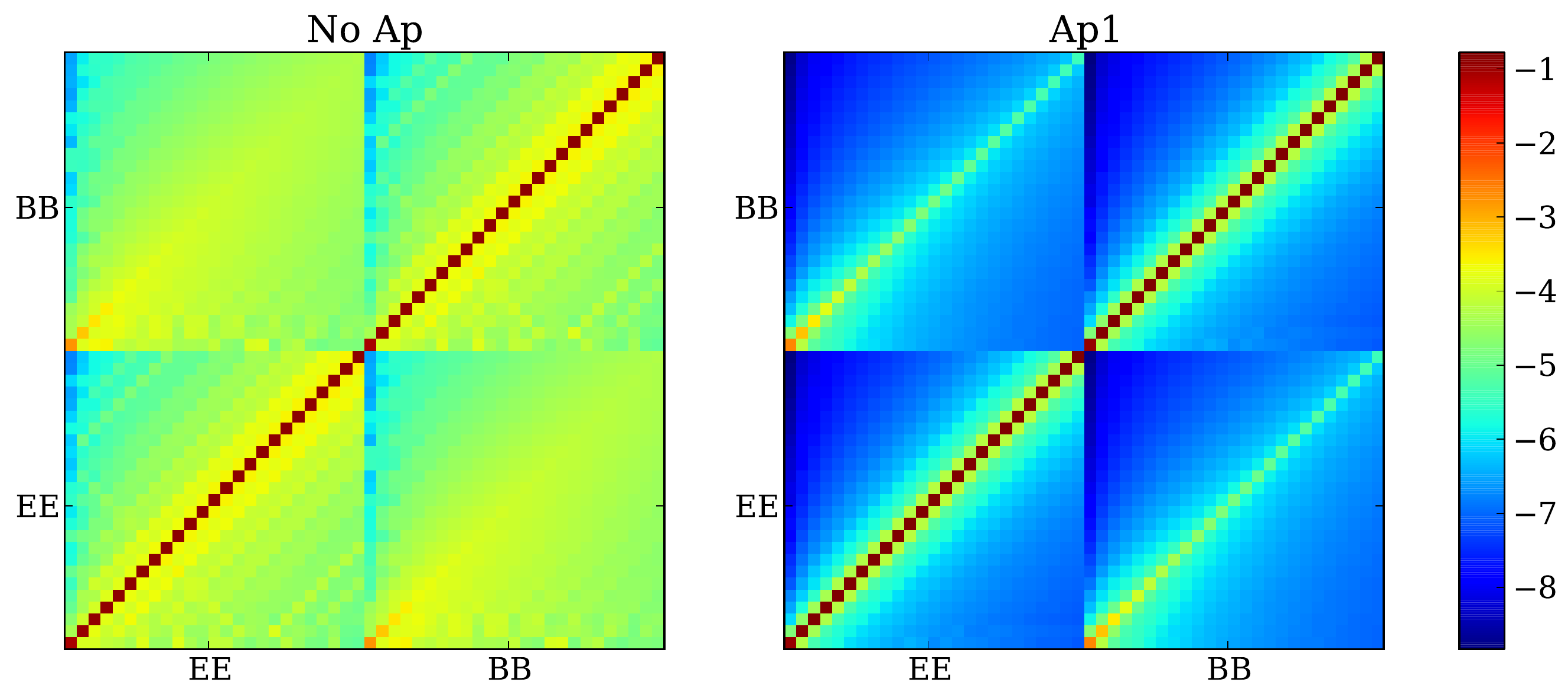}}  
     \caption{\small{The logarithm of the absolute value of the mixing matrix 
     for the composite mask (star and checkerboard). The left panel shows the mixing 
     matrix for the original mask with no apodisation, 
     whereas the right panels shows the same for an apodised mask with Ap1. 
     25 linear $\ell$-bins in $[245,8830]$ are considered here, 
     which corresponds to $n_\ell=20$.} }
    \label{figMixingMat}
  \end{center}
\end{figure*}

\fig\ref{figMixingMat} shows the mixing matrix for the composite mask (star and checkerboard).
The left panel shows the matrix for the original ones and zeros mask, while the right panel
shows the same for the mask apodised with Ap1.
Since the EB-EB part of the mixing matrix is independent of the EE and BB parts, it is not 
shown here and will not be used in any of the analysis. 
The mixing matrices are binned with $n_\ell=20$ to produce 25 approximately linearly spaced bins with
$\Delta\ell\approx n_\ell\times\Delta\ell_{\rm min}=360/{\rm rad}$.
The matrices are plotted in terms of the logarithm of the absolute value of their elements. 
As can be seen in this figure a smooth mask has a more diagonal mixing matrix and a smoother off-diagonal
behaviour. The importance of this property of the mixing matrix will become clear in the next sections.

\section{Measured power spectra}
\label{sec:PCLResults}

To test the mask modelling we use two sets of simulations: 
random realisations of Gaussian and lognormal shear fields.
The input power spectrum is identical for both cases and is based on a cold dark matter Universe 
with a dominant dark energy component, with cosmological parameters given in \tab\ref{tabCosmoPCL}. 

\begin{table}
\caption{\small{The fiducial cosmological parameters consistent with Planck 2013 results \citep[][]{Planck14}.
The normalization of the power spectrum, $\sigma_8$, is the standard
deviation of perturbations in a sphere of radius $8 h^{-1} {\rm Mpc}$ today. 
$\Omega_\mathrm{m}$, $\Omega_\Lambda$, and $\Omega_\mathrm{b}$ are the matter, 
dark energy and baryonic matter density parameters, respectively.
$w_0$ is the dark energy equation-of-state parameter, 
which is equal to the ratio of dark energy pressure to its density. 
The spectral index, $n_\mathrm{s}$, is the power of the initial power spectrum. 
The dimensionless Hubble parameter, $h$, characterizes the rate of expansion today.}}
\begin{center}
\begin{tabular}{ | c | c | c | c | c | c | c |  }
  \hline
 $\sigma_8$ & $\Omega_\mathrm{m}$ & $\Omega_\Lambda$ & $w_0$ & $n_\mathrm{s}$ & $h$ & $\Omega_\mathrm{b}$ \\
  \hline
   0.8  & 0.27  & 0.73  &  $-1.0$ &  0.96 & 0.72 & 0.045 \\
  \hline
\end{tabular}
\end{center}
\label{tabCosmoPCL}
\end{table}

The linear power spectrum is determined assuming 
a primordial power law power spectrum with 
\cite{EisensteinHu98} transfer function. 
Additionally, the halo fit formula of \cite{Smith03} 
is used for calculating the non-linear scales. 
We use a single redshift distribution of \cite{VanWaerbeke06} type with
$0.2<z<1.3$, a median redshift of 0.7, $\alpha=2$ and $\beta=1.5$.

All the simulations are originally made for a larger field 
($20^{\circ}\times20^{\circ}$, $2048\times 2048$ pixels), 
then a $10^{\circ}\times10^{\circ}$ field ($1024\times 1024$ pixels) is cut out of 
the middle to simulate the non-periodic nature of the Universe. 
100 random realisations are generated for each case in the analysis. 
After adding a Gaussian random shape noise with $\sigma_\epsilon=0.3$ to the shear fields, 
they are masked, and then zero-padded before the Fourier transform.
The zero-padding scheme used here changes the size of the fields to their original size, 
which means, doubling the size of the field on each side by adding zeros. 
The zero-padding ensures that a periodic boundary condition is not assumed for the field 
when the FFT is applied to it. Zero-padding the field more than this results in 
a computationally more expensive analysis while the result remains similar.
The FFT of a zero-padded field has a higher resolution, hence zero-padding is also a non-unique
form of interpolation between the Fourier modes. 
As a result the resolution of the Fourier transformed fields is $18/{\rm rad}$.

\begin{figure}
  \begin{center}
    \begin{tabular}{c}
      \resizebox{80mm}{!}{\includegraphics{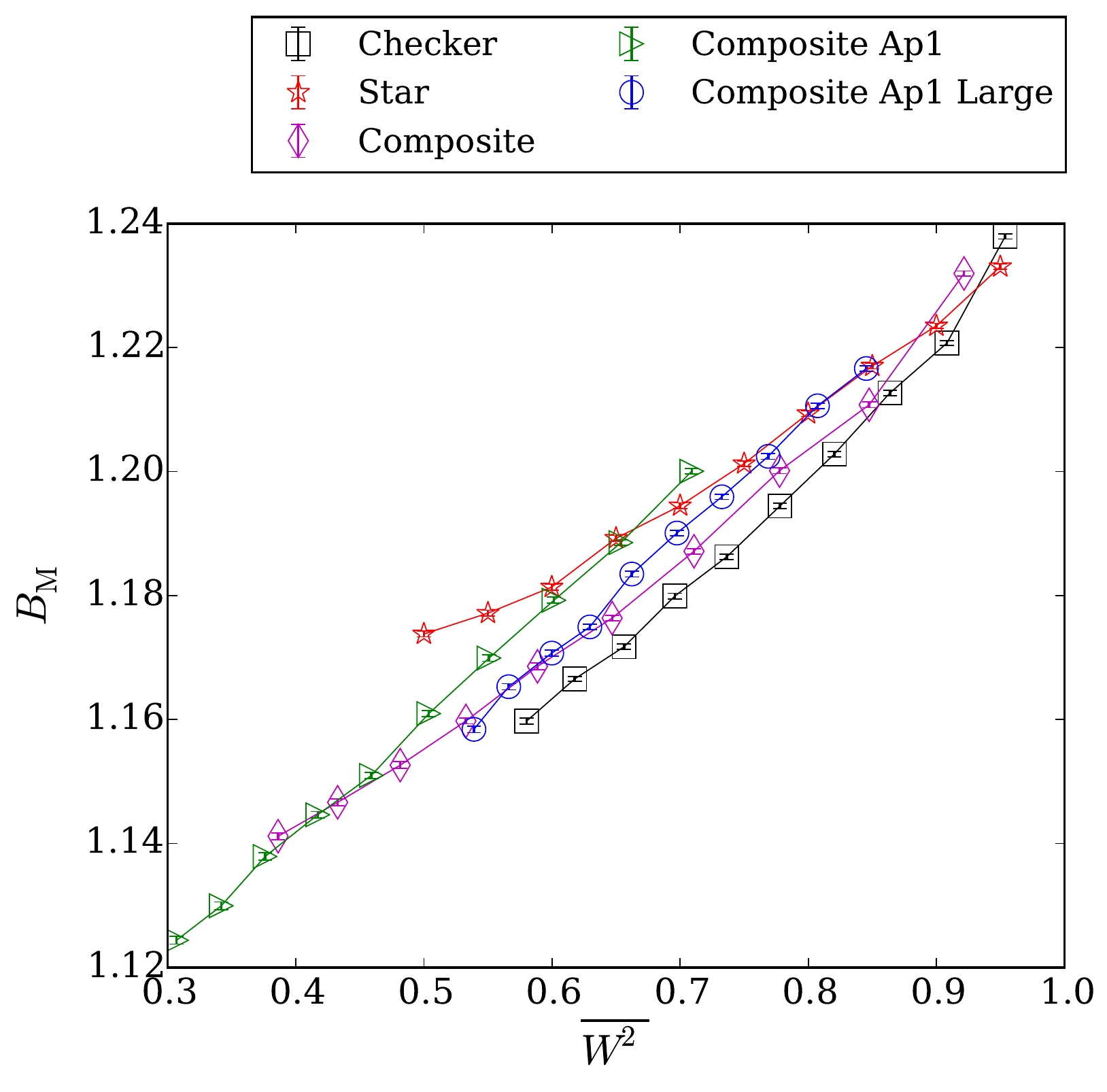}}     
    \end{tabular}
     \caption{\small{Constant multiplicative bias caused by inaccuracies in mask modelling on large scales. 
     The x-axis shows the mean value of the square of the mask, which is a measure of the masked area and the scaling
     effect of the mask. $B_{\rm M}$ is shown for five types of masks. 
     Star, checkerboard and composite masks are explained in \sect\ref{sec:Mask} 
     and the apodisation in \sect\ref{sec:Apodise}. 
     The blue circles show a modified composite mask which 
     consists of a coarser checkerboard pattern 
     (one rectangle for every four in the original checkerboard mask). 
     The solid lines simply connect the symbols to guide the eye.
     The error-bars show the variance of the mean between 
     the 100 simulated field and are smaller than the symbols for all cases.} }
    \label{figBias}
  \end{center}
\end{figure}

As we have seen in \sect\ref{sec:Mask} the mixing matrices calculated 
from pixelated masks are not accurate. 
The inaccuracy in mask modelling is more severe for
small $\ell$ which propagates to all scales (see \Eqt\ref{eq:M}). 
As a result, to first order a constant multiplicative bias needs to be corrected for. 
We measured this bias for different masks by taking the average ratio of the measured $\tC(\ell)$ to their theory 
value. This bias mainly depends on the mask and is shown in \fig\ref{figBias} as 
\begin{equation}
\label{eq:Bm}
 B_{\rm M}=\left\langle\frac{\tC^{\rm est}_\ell}{\tC^{\rm th}_\ell}\right\rangle\;,
\end{equation}
where the angled brackets mean the average of the ratio is taken over the 100 simulations and $\ell$-modes. 
We show $B_{\rm M}$ for star, checkerboard and three types of composite masks: 
composite without apodisation, with Ap1 and a composite with a larger checkerboard pattern apodised with Ap1. 
In \fig\ref{figMask} we showed the checkerboard mask we use in the main analysis. 
The larger checkerboard pattern is coarser with one large rectangle instead of every four 
smaller rectangles shown in \fig\ref{figMask}. 
The masks are varied so that the masked area changes. 
For the star mask this is done by keeping the size range of the stars as the original, 
but changing the percentage of star masks in each range. 
The checkerboard mask is varied by changing the number of dark pixels 
between the chips. For the composite masks both masks are varied simultaneously.
The change in the masked area is captured by $\overline{W^2}$, 
corresponding to the mean value of the square of the mask before zero-padding.
For a binary mask $\overline{W^2}$ is equal to the fraction of the sky which is unmasked.
Assuming no mode-mixing and zero-padding, applying a mask on the field results in the scaling of the power spectrum 
by $\overline{W^2}$. As can be seen in \fig\ref{figBias} the relation between $B_{\rm M}$ and $\overline{W^2}$ is 
roughly linear, however it is not universal and depends on the type of mask used. 
Generally, for heavier masking $B_{\rm M}$ decreases, this can be explained by looking at 
\fig\ref{fig:Wgg}. The first peak in the plots corresponds to the effect of zero-padding.
As we add more structures to the mask the relative significance of this peak decreases, since 
more masking produces power on other scales.
Fortunately, $B_{\rm M}$ is not very sensitive to the underlying power spectrum or the binning scheme.
For example in \fig\ref{figBias} we see that the scatter between the different realisations is very small,
which shows that the value of $B_{\rm M}$ is mostly sensitive to the mask and not the exact value of 
the underlying power spectrum.
$B_{\rm M}$ here is shown for the forward modelling case. 
A similar constant bias needs to be corrected for when $C^{\rm rec}_\ell$ is estimated, 
which has a similar behaviour and value.
This constant bias is only present when large scale masks are considered, therefore, \cite{Hikage11}
who only considered star masks with no apodisation, did not report it.

We tested several different methods to estimate the mixing matrix and concluded that the only way to 
systematically tackle this challenge is to find and correct for $B_{\rm M}$ using simulations. 
Note that while changing the integration scheme or the number of zero-pad pixels,
may alleviate this problem for certain masks it will not be applicable to others. 
For example, for a spherically symmetric mask, the best method to estimate the angular averages is 
to take averages over all the pixels (in Fourier space)
with the exact same distance to the middle of the field in Fourier space. 
Using this method results in a smaller number of points for each $\ell$-mode, 
which will in turn result in a very inaccurate mask modelling, for asymmetric masks. 
In conclusion, the method used in this paper is the most robust approach to mask modelling for a flat-sky analysis.
The results presented in this work have all been corrected with $B_{\rm M}$.
Nevertheless, we acknowledge that the need for this calibration makes this flat-sky FFT method less desirable
than an all-sky PCl method, which does not appear to produce this effect. 

Flat-sky PCl methods have been applied to data for cosmic microwave background temperature and polarization as well as cosmic infrared background anisotropy \citep{Planck11}. In those previous analysis a correction factor has been used to account for all the remaining data effects \citep[see for example][for POKER algorithm]{Ponthieu11}. This factor is estimated using simulations and thus, is in principle cosmology dependent. Applying this correction factor to the analysis can hide any residual masking effects. In the current work we treat this correction separately to understand its effects on flat-sky PCl estimators.

\begin{figure}
  \begin{center}
    \begin{tabular}{c}
      \resizebox{80mm}{!}{\includegraphics{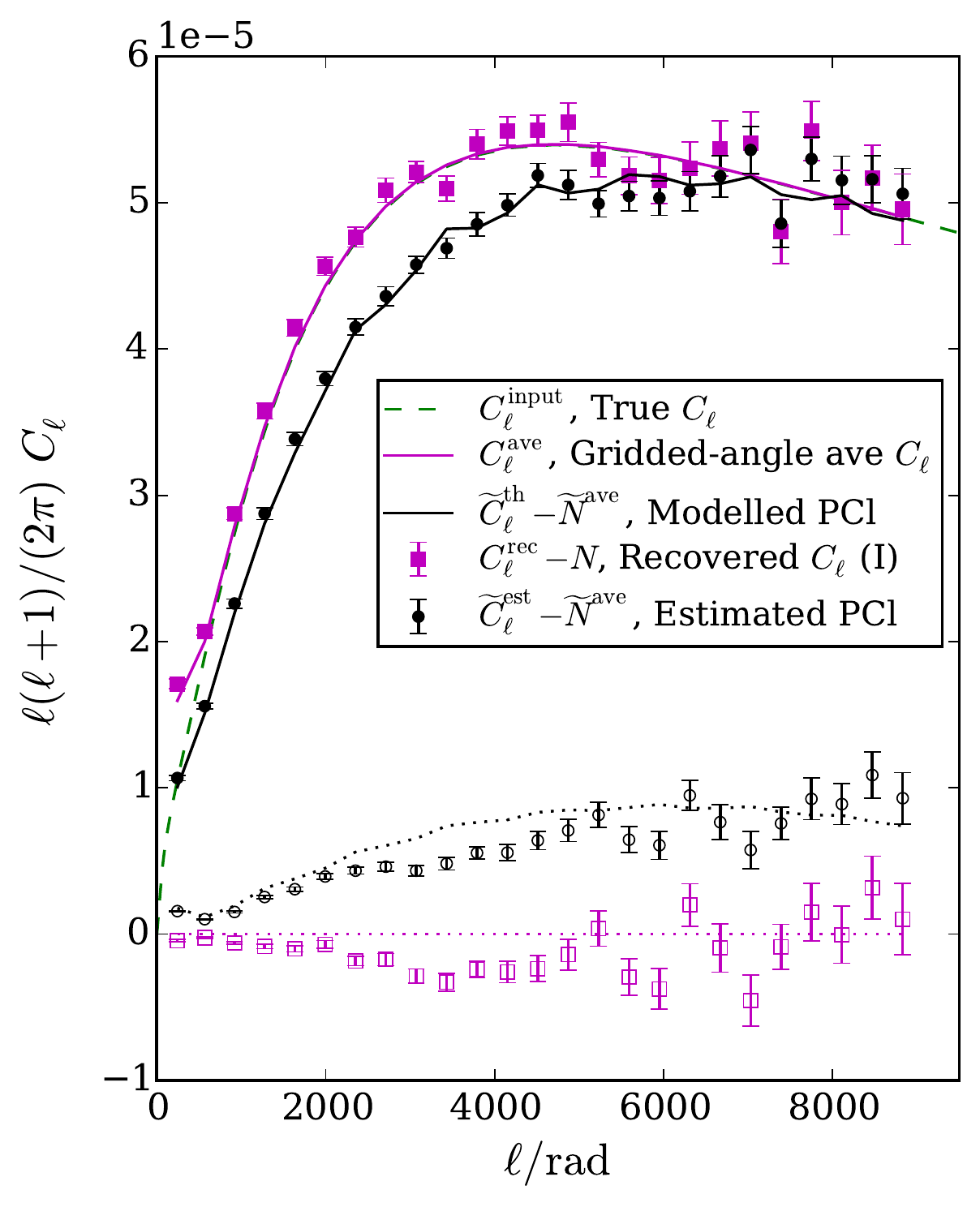}} 
     \end{tabular}
     \caption{\small{The estimated power spectra from the lognormal simulations and their expected values f
     rom theory for the composite mask with no apodisation. 20 of the smallest $\ell$-bins are merged 
     to make these power spectra ($n_\ell=20$, $\Delta\ell\approx360/{\rm rad}$).
     The curves show the expected theory values and the symbols show the estimated values from the simulations. The dashed green curve shows the input power spectrum from which the simulations are constructed. The magenta solid curve shows the input power spectrum after angle averaging.The magenta squares belong to the first recovery method defined in \Eqt\ref{eq:ClRecI}.
     The solid black curve and filled symbols show the theory and estimated E-mode PCls, while the dotted black curve and empty symbols show the theory and estimated B-mode PCls. The ellipticity noise contribution is subtracted from the results. The error-bars correspond to the field-to-field variations between the realisations of the shear fields. Similar plots for other masks are shown in \App\ref{appClPlots}.} }
    \label{figPCL}
  \end{center}
\end{figure}

\fig\ref{figPCL} shows the average estimated and theory PCls and Cls
of the lognormal fields for the composite mask, with $n_\ell=20$ 
(see \sect\ref{sec:Binning} for the details of the $\ell$-binning and the definition of $n_\ell$). 
The noise contribution is subtracted from the estimated (pseudo-)power spectra.
$C(\ell)^{\rm input}$ and $C(\ell)^{\rm ave}$ are shown 
in green dashed and magenta solid curve, respectively.
As discussed earlier, due to the pixelized nature of the fields the input power spectrum and its 
angle averaged version are not identical 
and show differences mostly at small Fourier modes, which is apparent in the figure as the small $\ell$ 
difference between the green dashed and magenta curves.
The recovered $C_\ell$ are shown as magenta squares. We use method I (see \Eqt\ref{eq:ClRecI}), which results 
in the closest recovery to the true power spectrum. Since the second recovery method (\Eqt\ref{eq:ClrecII}) does not provide any 
advantages (as discussed in \sect\ref{sec:Binning}) we will not use it in any of the following analysis.
The PCls shown in black are rescaled by a factor of $1/\overline{W^2}$, 
which enables us to see the mode-mixing effects of the mask.  
Furthermore, the noise contribution is subtracted from the PCls. 
The E-mode PCls are shown as a solid curve and filled circles while the B-mode PCLs are shown 
as a dotted curve and empty circles for theory and estimated values respectively.
This figure shows that the composite mask 
has a large effect on the PCls over a large range of Fourier modes.
Additionally, some of the power is moved to the B-modes. 
The theory and the estimated values of PCls are fairly consistent for the E-modes, 
which is not the case for the B-modes, since the B-modes have a smaller signal they are more
sensitive to inaccuracies in mask modelling, and hence will not be used for the parameter estimation in 
the next section. Similar plots can be seen in \App\ref{appClPlots} for the other masks.

\begin{figure}
  \begin{center}
    \begin{tabular}{c}
      \resizebox{80mm}{!}{\includegraphics{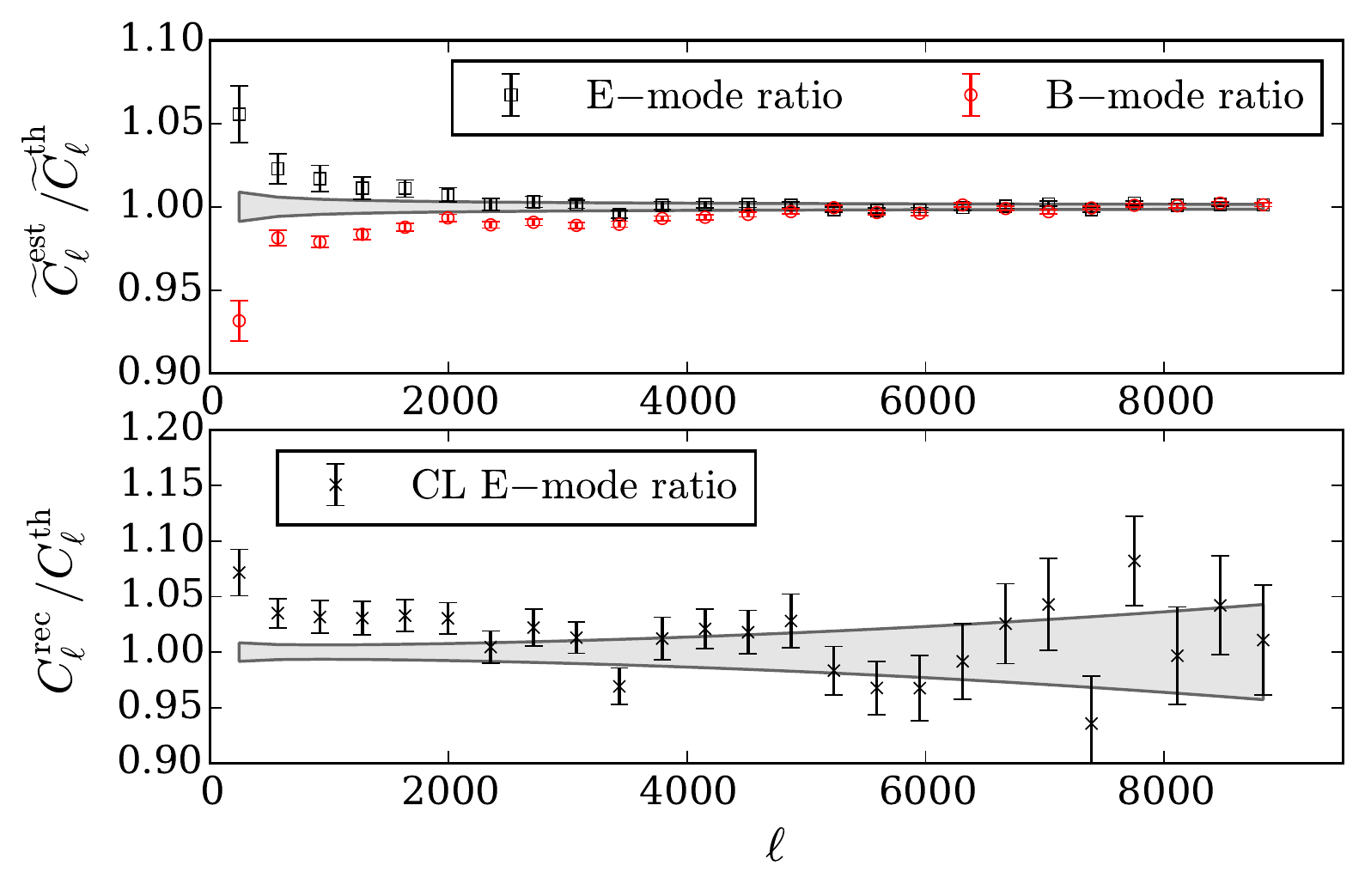}}    
    \end{tabular}
     \caption{\small{The ratio of estimated to theory power spectra for the composite mask 
     with $n_\ell=20$, $\Delta_\ell\approx 360/{\rm rad}$. 
     The top plot shows this ratio for the $\tC(\ell)$ and the bottom for $C(\ell)$.
     The noise contribution is not subtracted from $\tC(\ell)$,
     while it is subtracted from $C^{\rm rec}_\ell$, to recover the input power spectra.
     The black points show the ratio for the E-mode power spectra, whereas the red ones correspond to
     the B-modes. The recovered $C(\ell)$ is estimated from the first binning method, 
     for which the theory value of B-modes is zero, hence this ratio is not shown here. 
     The shaded area shows the expected cosmic variance. }}
    \label{figCLRatio}
  \end{center}
\end{figure}

Since the variance on the mean of 100 fields is very small, it is difficult 
to compare the theory to estimated $\tC(\ell)$ or $C(\ell)$ values in \fig\ref{figPCL}. 
Therefore, the relative power spectra are plotted in \fig\ref{figCLRatio}, for the composite mask. 
The grey areas show the cosmic variance for the simulated fields. 
The cosmic variance for $C(\ell)$ assuming a Gaussian distribution is estimated using,
\begin{equation}
 \label{eq:CosmicVar}
 \sigma^2_{\rm cosm,\;Cl}=\frac{2}{2\ell+1}\frac{1}{f_{\rm sky}\Delta\ell}(C(\ell)+N_\epsilon)^2\;,
\end{equation}
where $f_{\rm sky}$ is the fraction of the sky that is not covered by masks.
$\Delta\ell$ is the $\ell$-bin width and $N_\epsilon$ is the noise power.
For the 100 simulated fields of 100 square degrees each,
\begin{equation}
 f_{\rm sky} \simeq 0.24\; f_{\rm image}\;,
\end{equation}
where $f_{\rm image}$ is the effective fraction of each image not covered by the mask.
To find the cosmic variance for $\tC(\ell)$ we need to use the mixing matrix on the 
$\langle\Delta C(\ell)^2\rangle$. 
Doing so results in some off diagonal terms which we are not interested in, for the purposes of this section, 
since we only show the diagonal terms and their associated error-bars. 
The off-diagonal terms are incorporated in the analysis in \sect\ref{sec:FisherResults}.
The diagonal terms are
\begin{equation}
 \label{eq:CosmicVar2}
 \sigma^2_{\rm cosm,\;PCl}=\frac{2}{2\ell+1}\frac{1}{f_{\rm sky}\Delta\ell}(\tC(\ell)+\widetilde{N}_\epsilon)^2\;,
\end{equation}
which is the cosmic variance term for the PCls, where $\tC(\ell)$ is their expected value from theory.

The top plot  in \fig\ref{figCLRatio} shows the ratio of the estimated PCl to its theory value for E-modes and B-modes. 
Unlike in \fig\ref{figPCL} the noise contribution is not subtracted from the PCls,
since for the forward modelling Fisher analysis in the next section the PCls used contain noise.
As can be seen in the plots the low-$\ell$ ratios diverge from unity which shows an imperfect mask modelling 
at these scales. The E/B-modes show an almost anti-correlated behaviour on these scales which suggests that theory 
PCls does not account for all the mode-mixing and E/B leakage.
The bottom plot shows the ratio of $C(\ell)^{\rm rec}$ defined 
in \Eqt\eqref{eq:ClRecI} to its theory value which is equal to
the angle averaged input power spectrum, $C(\ell)^{\rm ave}$. 
Again here we see that the agreement between the theory and estimated values is better at scales above 
$\ell\sim2000$. 
Similar figures for other masks are provided in \App\ref{appClPlots}

\section{Error propagation}
\label{sec:PCLError}

The ultimate goal of a cosmic shear analysis is to constrain cosmological models and their parameters,
in a typical scenario.  
Here we use a Fisher analysis to put upper limits on the constraining power of
PCls. We use both $\tC(\ell)$ (forward modelling) and  $C(\ell)$ (backward modelling)
to compare the constraints and biases on model parameters.

\subsection{Fisher analysis: formalism}

The main purpose of this study is to determine the  accuracy of a PCl analysis and its limitations.
The ratio of the bias on a deduced model parameter 
to the errors associated with it can give us an indication of the accuracy of such analysis. 
Ideally, the estimators are unbiased, however, as we have seen
throughout this paper, there are sources of bias in a PCls analysis, 
originating from inaccuracies in mask modelling and binning effects. 

Formally, the Fisher matrix is defined as the ensemble average of second derivatives 
of the negative log-likelihood function at the maximum likelihood point,
\begin{equation}
\label{fisher}
F_{ij}\equiv \bigg\langle\frac{\partial^2 \mathfrak{L}}{\partial \phi_i \: \partial \phi_j} 
\bigg\rangle\; ,
\end{equation}
where $\phi_i$ are the parameters to be inferred. 

Assuming a Gaussian likelihood distribution, to calculate the Fisher matrix
we can use the following relation,
\begin{align}
\label{eq:Fisher}
F_{ij}=\langle \mathfrak{L}_{,ij} \rangle = 
\frac{1}{2} \mathrm{Tr}[\Cov^{-1}\:\Cov_{,i}\:\Cov^{-1}\:\Cov_{,j}+\Cov^{-1}\:M_{ij}]\;,
\end{align}
where $\Cov$ is the data covariance, $\Cov_{,i}$ is the derivative of $\Cov$ with respect to $\phi_i$,
$M_{ij}$ is a matrix composed of the derivatives of $\boldsymbol\mu$, the expected value of the data vector,
\begin{equation}
 M_{ij}=\boldsymbol{\mu}_{,i}\;\boldsymbol{\mu}_{,j}^\mathrm{t}
+\boldsymbol{\mu}_{,j}\:\boldsymbol{\mu}_{,i}^\mathrm{t}\;,
\end{equation}
where $\boldsymbol\mu^{\rm t}$ is the transpose of $\boldsymbol\mu$ 
(for the details of the derivation of \Eqt\ref{eq:Fisher} see \citealt{Tegmark97} for example).
The second term in \Eqt\eqref{eq:Fisher} is dominant for 
a survey with relatively large area as the covariance matrix 
is scaled inversely by the area \citep[see][where the exclusion of 
the first term is shown to have a negligible effect]{Asgari12,Eifler09}.
Hence we will only use the second term to calculate our Fisher matrices. 

The Fisher matrix can also be used to propagate the bias in the measured observables 
to the estimated parameters. \cite{Taylor07} showed that for a Gaussian distributed 
likelihood the linear bias for a parameter, $\phi_\mu$, given the bias in the observables, 
$\mathbf{x}$, is
\begin{equation}
\label{eq:Bias}
 B_\mu=\left(F^{-1}\right)_{\mu\nu} \mu_{i,\nu} \left(\Cov^{-1}\right)_{ij}
 \left(\mu_j-x_j\right) \;,
\end{equation}
where $\mu_j$ is the expected value of $x_j$ and Einstein summation rules apply
\citep[also see][]{Knox98,Kim04}. 

\subsection{Weak lensing covariance}
\label{sec:covariance}
To find the value for the Fisher matrix and the bias from Eqs.\ts\eqref{eq:Fisher} and \eqref{eq:Bias},
we need to find the covariance of $C(\ell)$ and $\tC(\ell)$. 
In this work we use lognormal shear fields as default 
and compare the final results with Gaussian shear fields. The covariance of a Gaussian
field has been calculated in the literature \citep[see][for example]{Kaiser98,JoachimiSE08}. 
In \App\ref{appLognormal} we show a general calculation 
for finding the moments of a lognormal field, which is then used to find the covariance of 
the shear power spectra.

A lognormal distribution provides a more realistic characterization of the convergence field,
$\kappa(\theta)$ \citep[see][and references therein]{HHS11}. 
Therefore, we can use the formalism in \App\ref{appLognormal} to estimate
the covariance of such a field. The equations in \App\ref{appLognormal} 
are derived for the moments of the density contrast, 
which has a minimum of minus one, unlike $\kappa(\theta)$. 
Hence we need to incorporate this difference to find the covariance of a lognormal $\kappa(\theta)$ field.
We can write the lognormal convergence field in terms of a Gaussian field,
\begin{equation}
 \kappa(\theta)=e^{n(\theta)}-\kappa_0\;,
\end{equation}
where $n(\theta)$ is a Gaussian random field and 
$\kappa_0$ is the absolute value of the minimum convergence \citep[see][]{HHS11}.
For this work the value assumed for $\kappa_0$ is 0.012 which corresponds to the value 
found by \cite{HHS11} for the Millennium simulation with source galaxy redshift of 0.76.
Using this definition instead of the one for $\dlg$ in \Eqt\eqref{eq:LogNormal} introduces 
extra constant coefficients in \Eqt\eqref{eq:Meanlgfinal}. The final result after 
applying these changes, are shown here.

The covariance matrix of the power spectrum of a lognormal convergence field 
can be written in terms of the sum of the covariance of a Gaussian field and a purely 
lognormal term,
\begin{align}
\label{EqCovTot}
 \Cov^{\rm tot}(\ell,\ell')&\equiv\Cov^{\rm ln}(\ell,\ell')+\Cov^{\rm G}(\ell,\ell')\\ \nonumber
 &=\langle \hat C(\ell) \hat C(\ellp) \rangle-C(\ell) C(\ellp)\;,
\end{align}
where $C(\ell)$ is the expected value of $\hat C(\ell)$.  $\Cov^{\rm G}(\ell,\ell')$ is given in \cite{Kaiser98} as,
\begin{align}
\label{eq:CovGauss}
\Cov^{\rm G}(\ell,\ell')=\frac{4\pi}{A\ell\Delta\ell}(C(\ell)+N_\epsilon)^2\delta_{\ell\ell'}\;,
\end{align}
where $\delta_{\ell\ell'}$ is the Kronecker delta, which makes this covariance diagonal.
$N_\epsilon$ is the noise power spectrum given in \Eqt\eqref{eq:NoisePower}. 
$\Delta\ell$ is the width of the $\ell$-bin and $A$ is the area of the field.
Note that this a model covariance matrix which assumes a simple survey geometry. 

The purely lognormal term, $\Cov^{\rm ln}(\ell,\ell')$, can be calculated using the 
purely lognormal terms of the 4th order lognormal moments.
We put $k_1=\ell$, $k_2=-\ell$, $k_3=\ellp$ and $k_4=-\ellp$ in \Eqt\eqref{eq:4thMoment},
following \cite{HHS11} we ignore all terms but $IV$, which we simplify to find the desired relation,
\begin{align}
 \Cov^{\rm ln}(\ell,\ell')&\simeq
 \frac{1}{A^2\kappa_0^2}\int \frac{\d \varphi_\ell}{2\pi}\int \frac{\d \varphi_\ellp}{2\pi}
 \langle\tkappa(\ell)\tkappa(-\ell)\tkappa(\ell')\tkappa(-\ell')\rangle_{IV}\\ \nonumber 
 &= \frac{1}{A\kappa_0^2}\big\{ 2[C^2(\ell) C(\ellp)+C^2(\ellp) C(\ell)]\\ \nonumber
 &+[C(\ell)+C(\ellp)]^2 \int \frac{\d \varphi_{\ell\ellp}}{2\pi}[C(|\lb-\lb'|)+C(|\lb+\lb'|)]\big\}\;,
\end{align}
where $\varphi_{\ell\ellp} = \varphi_{\ell}-\varphi_{\ellp}$, is the angle between $\lb$ and $\lb'$.

To find the binned covariance matrix, we first calculate the covariance matrix for the smallest binning 
and then apply the binning matrix to it,
\begin{equation}
\label{eq:covBack}
 \Cov^{\rm tot}_{\rm b}=\B\Cov^{\rm tot}\B^{\rm t}\;,
\end{equation}
where $\B$ is the binning matrix defined in \sect\ref{sec:Binning}. 
The covariance matrix for a binned $\tC(\ell)$ is then,
\begin{equation}
\label{eq:CovPCL}
 \widetilde\Cov^{\rm tot}_{\rm b}=\B\MM\Cov^{\rm tot}\MM^{\rm t}\B^{\rm t}\;.
\end{equation}
The cosmic variance term in \Eqt\eqref{eq:CosmicVar} is basically the same as the diagonal terms
in \Eqt\eqref{eq:CovGauss} with $2\ell+1\approx2\ell$. 
The area in \Eqt\eqref{eq:CovGauss} is the area of the field before zero padding, hence here we ignored the 
masking effect. But in \Eqt\eqref{eq:CosmicVar} we use the effective area of the field.
The reason behind this difference is that we use the unmasked covariance on 
the right hand side of \Eqt\eqref{eq:CovPCL} to find the masked one, which accounts for the loss of area.
Note that the covariance in \Eqt\eqref{eq:CovGauss} is an analytic estimate for a Gaussian field which is
simply connected and when all relevant angles are smaller than the extent of the field \citep[see][]{JoachimiSE08}. 
In the next section we use the area before masking to measure the covariance of the recovered Cl. 
This is not an accurate representation of this covariance as we are assuming all the lost information
due to masking, is recovered. Scaling this covariance with the effective area is not a fair representation
either. However, in \tab\ref{tab:area} we provide effective area scaling factors which can be used to rescale 
the covariance values in the following section.

Here we limit our theory and simulation comparison to power spectra and leave a covariance matrix comparison to future work. For our finest binning we have 500 $\ell$-modes. To get an unbiased estimate of the inverse covariance, at least 625 simulations are needed according to \cite{Anderson} or $\sim 4000$ for a more accurate estimate \citep[better than 5$\%$, see][]{Hartlap07}.

\begin{table}
\caption{\small{The area scaling due to mask. The values in the table correspond to the ratio of the
field area after to its area before masking ($f_{\rm sky}/A$).}}
\begin{center}
\begin{tabular}{ | c | c | c | c | c |  }
 No Ap & Ap1  & Ap2  & Ap3   &  \\
  \hline
 1.00  & 1.00 & 1.00 & 0.99  & No Mask\\
 0.90  & 0.90 & 0.90 & 0.89  & Star\\
 0.86  & 0.70 & 0.49 & 0.17  & checkerboard\\
 0.78  & 0.63 & 0.43 & 0.15  & composite\\
  \hline
\end{tabular}
\end{center}
\label{tab:area}
\end{table}

\subsection{Fisher analysis: results}
\label{sec:FisherResults}

In this section we investigate the significance of the possible biases in the estimated cosmological parameter 
due to a PCl analysis, and compare that to the errors on the estimations. 
We calculate the bias on the parameter estimation from \Eqt\eqref{eq:Bias} for each 
simulated field,  by comparing the expected value of the power spectra to their observed value. 
The mean and standard deviation of the bias are then measured from 
100 realisations of the shear field for each set of the simulations. 
The error on the other hand is calculated analytically using a Fisher analysis (see \Eqt\ref{eq:Fisher}) and 
a theoretical covariance matrix for the power spectra (see \sect\ref{sec:covariance})
for a single 100 deg$^2$ field. The error can be scaled by $1/10$ to obtain the error for 
a 10,000 deg$^2$ field.

We limit our study to two free parameters, $\sigma_8$ and $\Omega_{\rm m}$,
to demonstrate the validity of the PCl method for weak gravitational lensing. 
The number of free parameters does not change the main results. 

We use a fixed $\ell$-mode range ($\sim$ (18-9000)rad$^{-1}$) but in reality a redshift-dependent maximum 
$\ell$-mode should be used in order to create a consistent $k$-mode selection. 
This relation is approximately $\ell_{\rm max}=k_{\rm max}r(z)$, where $r(z)$ is the comoving distance. 
We choose a maximum $\ell$-mode of $\sim9000$ which corresponds 
to the largest angular mode that would be probed at the largest comoving distance for a $k_{\rm max}\simeq 1$.
Some surveys might truncate at smaller $\ell$, especially due to uncertainty in nonlinear modelling.
In addition the S/N of a power spectrum estimator is relatively flat between $\ell=1000$ and $\ell=10000$
\citep[see for example][]{Sato13}. 
Nevertheless, here we want to demonstrate the constraints from the full $\ell$-range,
which can be relevant for baryonic physics.

Recall that in forward modelling we apply the mixing matrix on the theory power spectra, 
while in the backward modelling we instead correct the masking effects 
by applying the inverse mixing matrix to the observed PCls. 
We only use the E-mode power spectra for this analysis.

\begin{figure*}
  \begin{center}
    \begin{tabular}{c}
      \resizebox{150mm}{!}{\includegraphics{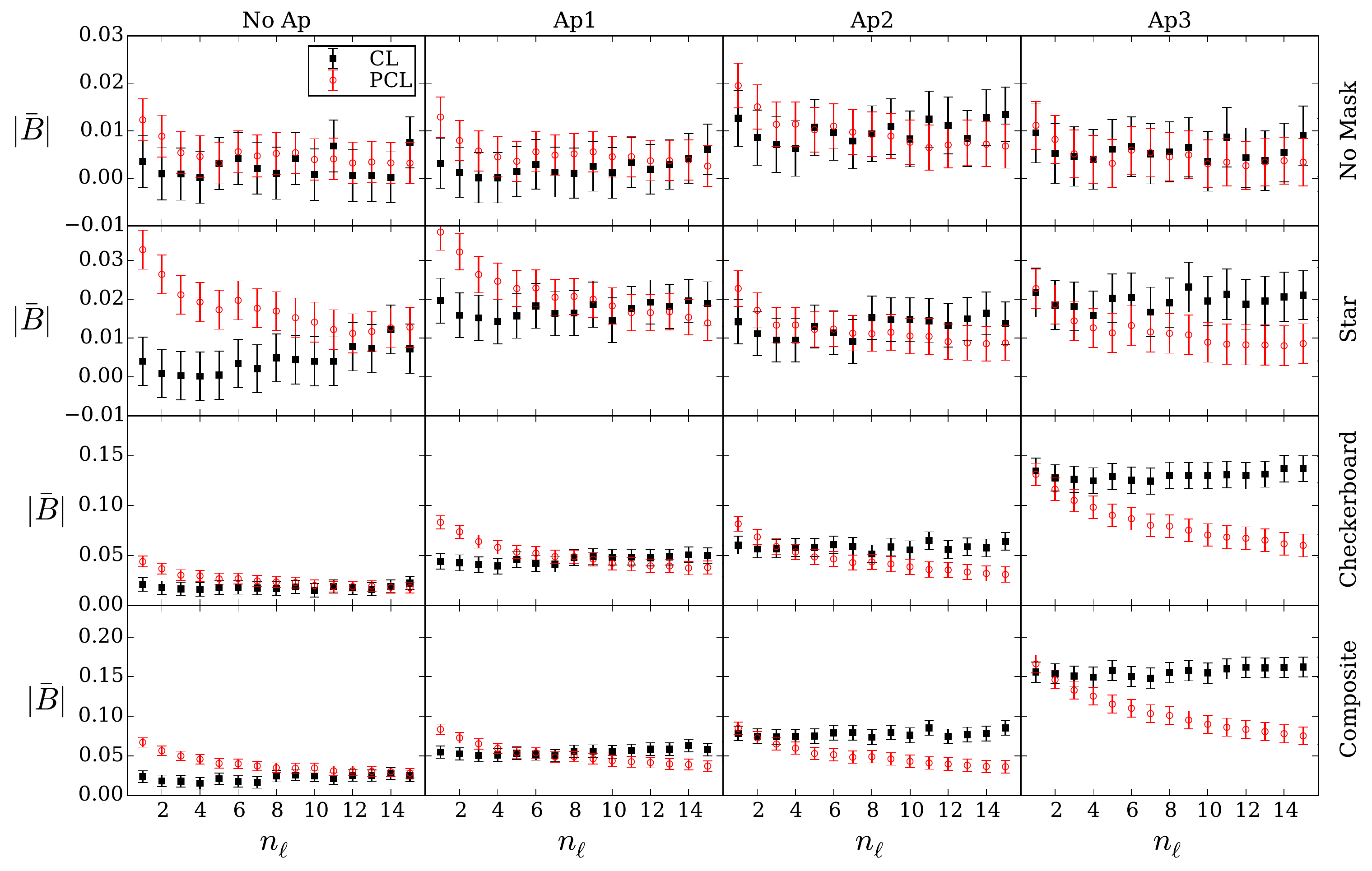}}  
    \end{tabular}
     \caption{\small{The average bias on the estimated parameter, $\sigma_8$, with respect to 
     the bin size. $n_\ell$ is the number of original $\ell$ bins which are combined to make the wider bins.
     All the other cosmological parameters are fixed to their fiducial values in \tab\ref{tabCosmoPCL}, aside from
     $\Omega_{\rm m}$ which is marginalized over. The red empty circles denote the forward modelling scheme, 
     where the PCls are the observables and the theory mixing matrix is applied to the theory power spectra, 
     while the black full squares show the values for the backward modelling, 
     where the mask correction is applied to the data 
     instead of the theory power spectra. Each row belong to a different mask 
     and each column to a different apodisation scheme (see \sect\ref{sec:Apodise}).
     A Fisher analysis is used for estimating the bias on the estimated parameter. 
     The errorbars show the error on the mean estimated from the field-to-field variance between the 
     100 simulated fields. The lognormal simulations are used here.} }
    \label{fig:Bias}
  \end{center}
\end{figure*}

\begin{figure*}
  \begin{center}
    \begin{tabular}{c}
      \resizebox{150mm}{!}{\includegraphics{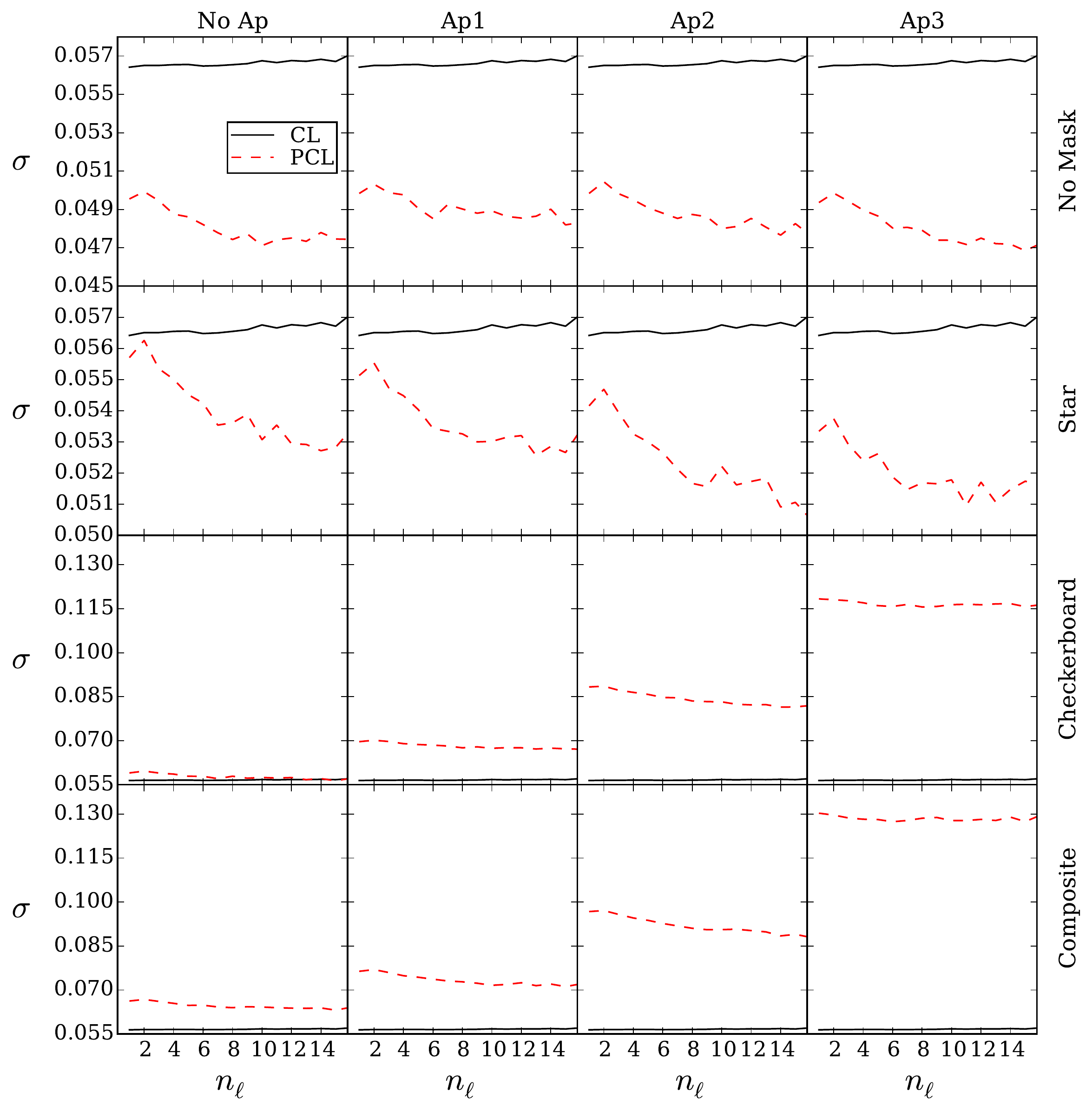}}  
    \end{tabular}
     \caption{\small{The one sigma error on the estimated parameter, $\sigma_8$, with respect to 
     the bin size. $n_\ell$ is the number of original $\ell$ bins which are combined to make the wider bins.
     All the other cosmological parameters are fixed to their fiducial values in \tab\ref{tabCosmoPCL}, aside from
     $\Omega_{\rm m}$ which is marginalized over. The red dashed line belongs the forward modelling scheme, 
     where the PCls are the observables and the theory mixing matrix is applied to the theory power spectra, 
     while the black solid line shows the values for the backward modelling, 
     where the mask correction is applied to the data 
     instead of the theory power spectra. 
     Each row belong to a different mask 
     and each column to a different apodisation scheme (see \sect\ref{sec:Apodise}).
     Note that the black solid line is remains constant between the different panels, since it is unaffected by the mask. 
     A Fisher analysis is used for estimating the error on the estimated parameter. } }
    \label{fig:Error}
  \end{center}
\end{figure*}

\begin{figure*}
  \begin{center}
    \begin{tabular}{c}
      \resizebox{150mm}{!}{\includegraphics{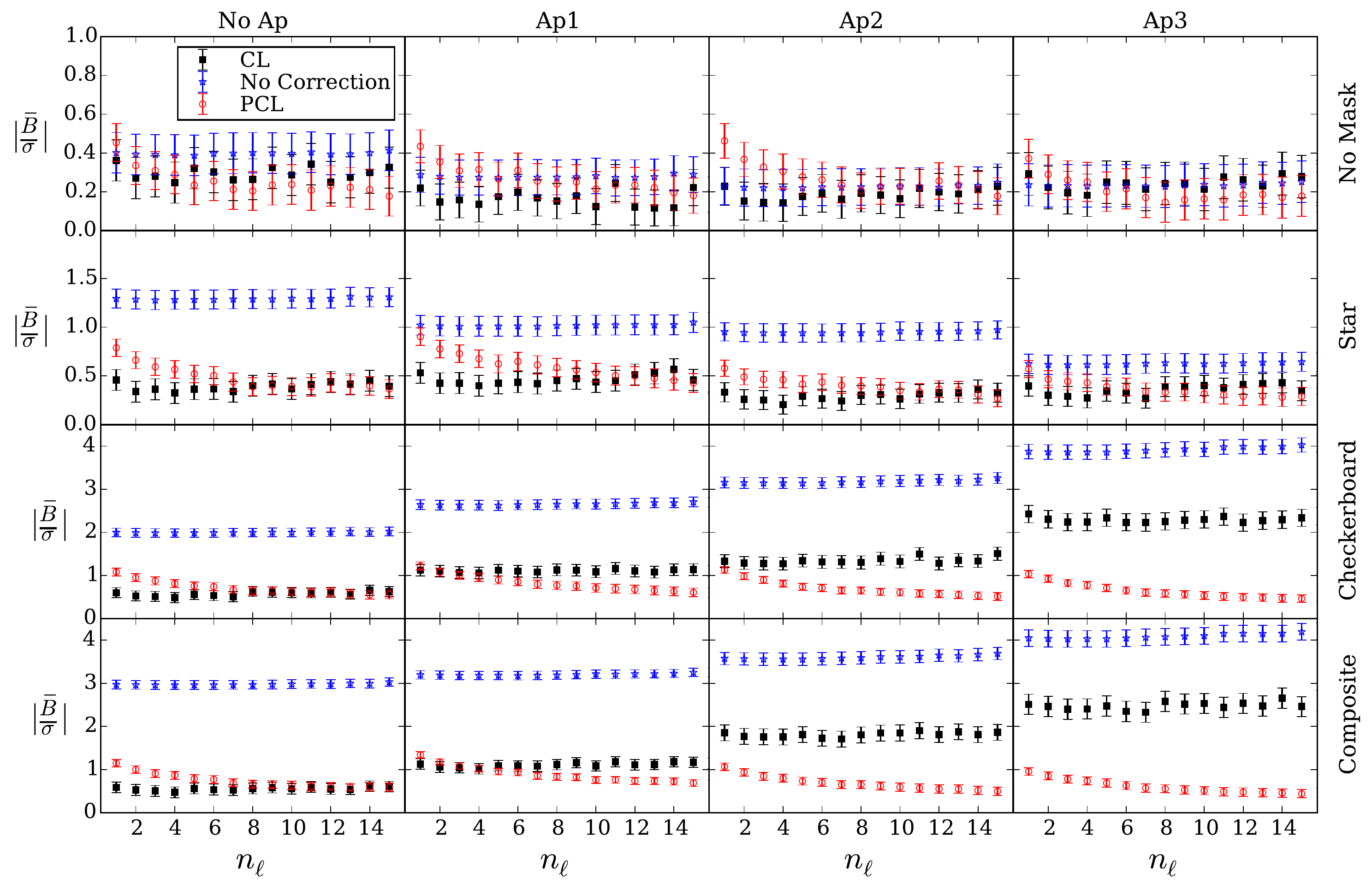}}  
    \end{tabular}
     \caption{\small{The ratio of the average bias to the one sigma error on the estimated parameter, 
     $\sigma_8$, with respect to the bin size. 
     $n_\ell$ is the number of original $\ell$ bins which are combined to make the wider bins.
     All the other cosmological parameters are fixed to their fiducial values in \tab\ref{tabCosmoPCL}, aside from
     $\Omega_{\rm m}$ which is marginalized over. The red empty circles belongs the forward modelling scheme, 
     where the PCls are the observables and the theory mixing matrix is applied to the theory power spectra, 
     while the black solid line shows the values for the backward modelling, 
     where the mask correction is applied to the data 
     instead of the theory power spectra. The blue stars show the ratio for the case where 
     no mask correction is applied to either the theory or the observed power spectra 
     aside from a multiplicative area correction factor. The blue symbols show 
     the level of importance of the mask correction for different masks.
     Each row belong to a different mask and each column to a different apodisation scheme (see \sect\ref{sec:Apodise}).
     A Fisher analysis is used for estimating the bias and error on the estimated parameter. 
     The errorbars show the error on the mean estimated from the field-to-field variance between the 
     100 simulated fields. The Gaussian simulations are used here.} }
    \label{fig:BiasToError}
  \end{center}
\end{figure*}

Figures\thinspace\ref{fig:Bias}, \ref{fig:Error} 
and \ref{fig:BiasToError} show the absolute value of the mean
bias, $|\bar{B}|$, the one sigma error and their ratio 
for $\sigma_8$, while $\Omega_{\rm m}$ is marginalized over,
respectively. The results for $\Omega_{\rm m}$ are not 
shown here as they closely follow the ones for $\sigma_8$.
Each row belongs to a different mask which 
is named at the right hand side of the row and each column 
to a different apodisation scheme named at the top of the column. 
All three figures show the results for all mask and apodisation combinations. 
For the description and definition of the masks and apodisation cases 
see \sect\ref{sec:Mask} and \sect\ref{sec:Apodise}. 
The x-axis in these plots, $n_\ell$, is the number of 
the smallest bins which are combined to make a wider bin, 
therefore, the width of the corresponding bin 
is $n_{\ell}$ times larger than the smallest bin. 
The smallest bin width is $~18/{\rm rad}$. 
The red empty circles and dashed lines belong to forward modelling, 
while the black full squares and solid lines show the backward modelling values. 
In \fig\ref{fig:BiasToError} blue stars show the case where 
no mask correction, other than a constant area factor, has been applied. 
This is shown for a comparison with the corrected versions. 
The $\sigma$ for the blue stars is the same as the backward modelling case. 
Note that the $\sigma$ in \fig\ref{fig:Error} 
is an analytic calculation which is noise free, 
however, the bias calculation changes for each field and hence there 
is a scatter between them which is captured by the 
error-bars in Figs\thinspace\ref{fig:Bias} and \ref{fig:BiasToError}.

By studying the three Figures\thinspace\ref{fig:Bias}, 
\ref{fig:Error} and \ref{fig:BiasToError}, 
we can see that for the cases with no 
apodisation the best method to use is the backward modelling, 
as the bias is consistently the lowest as well as the bias to error ratio. 
Note that in \fig\ref{fig:Error} the black solid line remains 
constant over all the panels as the backward modelling covariance 
has no information about the mask (see \Eqt\ref{eq:covBack} and the discussion that follows).
On the contrary, the covariance for forward modelling depends on the mask.

The forward modelling bias decreases as $n_{\ell}$ 
and hence the bin width increases in agreement with
 \cite{AsgariSchneider15} who showed that narrower 
 band power spectra are generally more biased. 
 We apply a binning matrix (see \Eqt\ref{eq:Binning}) 
 to our modelling to minimize this effect. 
 The binning works better for the backward modelling as can be seen 
 from the approximately flat behaviour of the bias with respect to $n_{\ell}$
 in \fig\ref{fig:Bias}.

A comparison of different masks in \fig\ref{fig:Bias}
shows that the large scale mask is generally more difficult to model and results in 
a larger bias. This effect is more pronounced for the composite mask where 
all the scales are affected. 
Note that in all cases in this work zero padding is present, 
which affects the small $\ell$-modes, 
and propagates through to all modes in the mixing matrix estimation, 
this was corrected for earlier by a multiplicative 
factor which also depends on the other properties of the mask (see \fig\ref{figBias}).
Apodising the mask increases the area covered by the mask, which in turn results in a more
biased estimate for $\sigma_8$. In \fig\ref{fig:Bias} we see that 
the apodisation affects the backward modelling more than the forward case, 
especially when the checkerboard mask is present.
Ap3 which has the largest kernel out of the 3 smoothing schemes, 
has a drastic effect on both modelling schemes for the
smallest $\ell$-bins, but the PCl values recover after binning. 
Furthermore, in \fig\ref{fig:BiasToError} we see that the 
ratio of bias to error is hardly affected for forward modelling 
and seems to improve in contrast to backward modelling. 
This can be explained by looking at \fig\ref{fig:Error} 
where we see that the one sigma error on $\sigma_8$ is adjusted 
in the forward modelling case by the mixing matrix which allows for a lower bias to error ratio.

The general conclusion from inspecting Figures\thinspace\ref{fig:Bias}, \ref{fig:Error} and \ref{fig:BiasToError}
is that if the data is masked in a binary manner (ones and zeros mask), 
which is the "no Ap" case, a backward modelling where
the recovered $C(\ell)$ are measured provides a better method, 
while for a non-binary mask or other effects that can mimic such masks, 
the forward modelling provides a better choice. 
For example, the inverse variance weight on shape measurements 
which has a multiplicative effect on the measured ellipticities 
can be interpreted as an apodised mask \citep[see][for the definition of the inverse variance weights]{Miller13}. 
If the combination of these weights and the already present masks  form a uniform structure 
in the images they will resemble the apodised checkerboard mask and hence we expect them to behave similarly. 

\begin{figure*}
  \begin{center}
    \begin{tabular}{c}
      \resizebox{170mm}{!}{\includegraphics{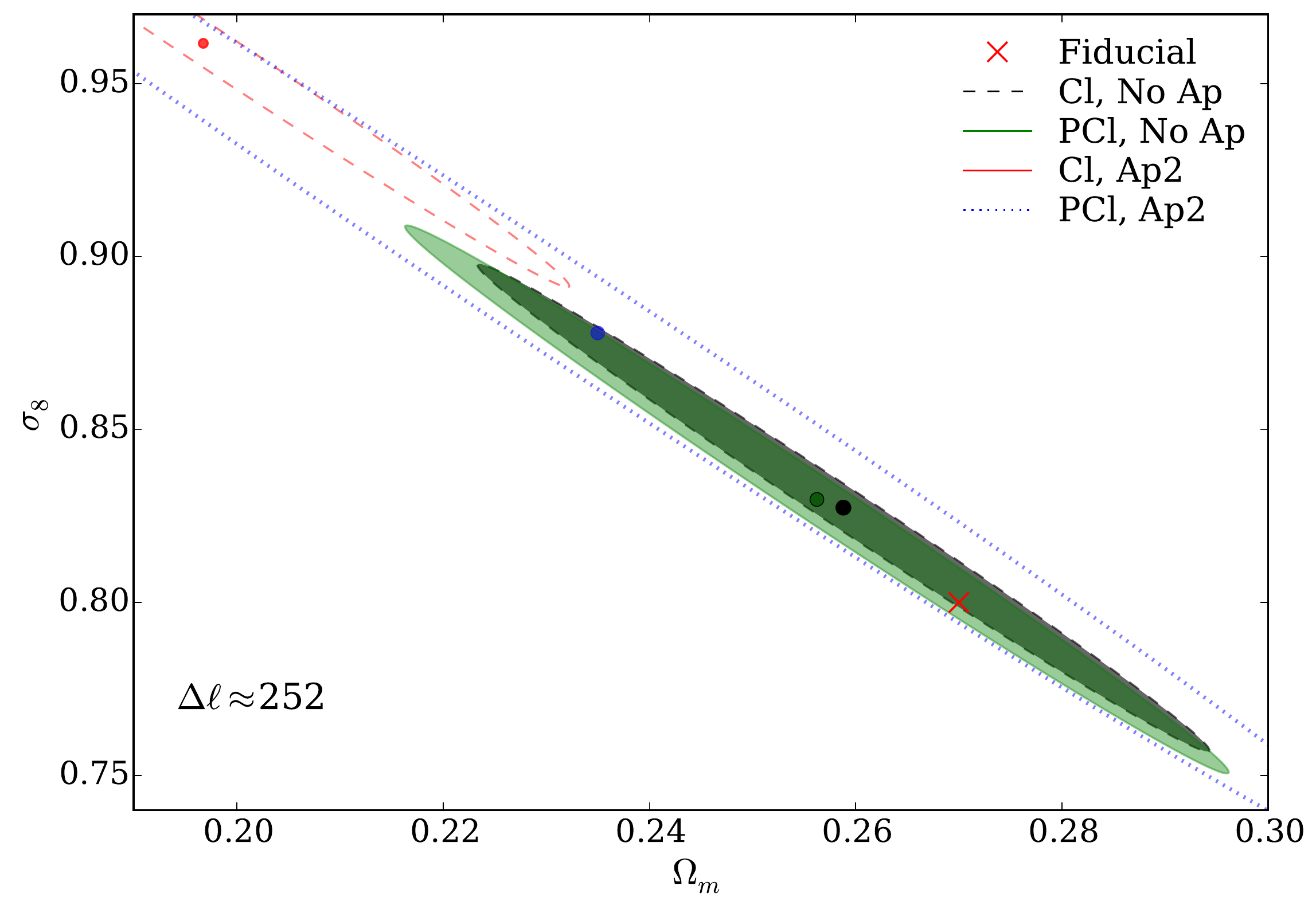}}  
    \end{tabular}
     \caption{\small{The $95\%$ Fisher contours for four different cases. The fiducial value of $\sigma_8$
     and $\Omega_{\rm m}$ are marked by a red cross. The ellipses are shifted according to the mean bias over 
     100 lognormal shear field realisations. The composite mask is considered here, with and without apodisation. 
     The empty contours refer to the two apodised cases using Ap2 (see \tab\ref{tabAp}),
     while the filled ones belong to the non-apodised masks.     
     The filled dashed (grey) and the empty dashed (red) contours belong to backward modelling, where
     the power spectrum is recovered, whereas the the green filled solid and the blue dashed contours show the results for forward modelling. 
     The power spectra are binned with $n_\ell=14$ which corresponds 
     to an approximately linear binning with $\Delta\ell\approx252$. The size of these contours correspond to a survey
     of 100 deg$^2$ and $n_{\rm gal}=30/{\rm arcmin}^2$.} }
    \label{fig:Fisher}
  \end{center}
\end{figure*}

\fig\ref{fig:Fisher} summarises the main conclusions in this section. It shows the Fisher constrains and the linear bias
in the $\Omega_{\rm m}$-$\sigma_8$ plane. The ellipses show the $95\%$ confidence regions and they are shifted from the 
fiducial position according to their bias value. The fiducial position of the parameters is shown as a red x. 
The results are shown for the composite mask with no apodisation and "Ap2" for backward (recovering $C_\ell$) 
and forward modelling (applying the mixing matrix to the theory).
The non-apodised cases are shaded, the dashed black one shows the backward modelling case, while the solid green one
belongs to forward modelling. The ellipse sizes do not change for backward modelling as we saw in \fig\ref{fig:Error}. 
The empty ellipses belong to 
the apodised case with "Ap2" (see \tab\ref{tabAp}), where the dotted blue refers to 
forward and the dashed red to backward modelling.
The results are shown for the binning case with $n_\ell=14$, $\Delta \ell\approx 252$. 
Aside from the backward modelling for the apodised mask, 
all the biases are within the $95\%$ contours. 
The apodisation changes the size of the forward modelling ellipse, since apodising the composite mask
results in significant loss of area which is captured by the mixing matrix (see \Eqt\ref{eq:CovPCL}).
However, the analytical covariance for the backward modelling in \Eqt\eqref{eq:covBack} 
assumes a simplistic survey geometry and hence is not exact, especially for a heavily masked region. 
Consequently, the backward modelling shows large bias to error ratios
for an apodised mask with regular feature, such as the checkerboard case (see \fig\ref{figMask}).

\section{Conclusion}

Pseudo-Cl analysis is a method that models the effects of masks on the Fourier transform of a field. It provides a crude but fast, FFT-based direct measurement of the power spectrum in the presence of masks, but may require calibration to simulated data. In this paper we have applied the flat-sky PCl approximation on simulated shear fields to investigate the accuracy and potential biases in this method for weak gravitational lensing analysis. This is particularly interesting for both current surveys, where the flat-sky approximation is used, and for future large-scale surveys such as Euclid\footnote{http://sci.esa.int/euclid/} and LSST\footnote{http://www.lsst.org/lsst/} where rapid methods may be useful.
However, we note that, given the substantial need for calibration, running a flat-sky analysis on data that covers large parts of the sky is not advantageous as the gain in computational speed would at best be minimal. It remains potentially competitive on small survey patches, e.g. early Euclid data.
Although, flat-sky PCl has been used for cosmic microwave background analysis, it has never been tested to the extent that is done here.  Here we show, for the first time, the effects of incomplete mask modelling in a flat-sky implementation of the PCl method on the estimated cosmological parameters.

Masking introduces mixing of Fourier modes in the shear field, and when the masked field is decomposed into E-convergence and B-curl modes there is mixing between the E/B-modes. In order to forward-model this effect, or recover the all-sky power from a masked field by deconvolution, the mask mixing matrix has to be modelled to a high accuracy. We have shown the need to carefully model the mask mixing matrix, taking into consideration the numerical estimation of integrals on a pixelized field. To investigate the effects of masking, we applied large-area masks, corresponding to the limits of a survey, and small-area sub-masks which would model the presence of star masks in a field, and a checkerboard pattern to model the effects of field-of-view boundaries for a mosaicked survey observing strategy. In a previous study, \cite{Hikage11} performed an analysis of simulated shear fields with full and flat-sky PCls. However they only considered small-scale star masks, with periodic boundary conditions, and did not propagate the errors to the measured parameters. Hence the analysis provided here complements and goes beyond \cite{Hikage11}. To maintain realism and avoid periodic boundary conditions, all the fields used in this work  were cut out of larger fields for both sets of Gaussian and lognormal simulated fields. As a result, when analysing the fields we first zero-padded them to their original size. Consequently, a large scale mask was present for all masks in this analysis.

We find that for a flat-sky implementation of the Pseudo-Cl weak lensing power spectrum analysis an overall, constant calibration correction for large-scale masks is required due to the sparsely-sampled low-l modes, and rapid oscillations of all modes of the mask power spectrum (see \fig\ref{fig:Wgg}). We have shown that this calibration bias is insensitive to the input power spectrum. For small-area star masks, the forward modelling and all-sky recovery both work well, resulting in slight biases in the lensing power spectra at a few percent for $l<2000$. There is also a slight improvement with modest apodisation of the mask. For the checkerboard mask, again there is good modelling and all-sky recovery on all scales above $l=2000$, but the large-angle bias is slightly worse, again a few percent, which apodisation makes worse. While apodisation suppresses small-scale mask power, it does not help with the large-scale power and rapidly oscillating power which lead to biases. Indeed, apodisation introduces significant biases into the full-sky retrieval, while notably  increasing both bias and errors in forward modelling due to the loss of effective sky area.

While investigating the recovered all-sky power spectrum, we found that the choice of binning of the mixing matrix made a significant difference. If we bin the angular wavenumber in wider bins the mask mixing matrix becomes more diagonal which results in an overall scaling of the masked power spectra. This arises due to the loss of the fine-structure in the mask power spectrum by l-binning. However, if we evaluate the mixing matrix per-$\ell$, deconvolve and then re-bin, we preserve the fine-structure of the mask power, and the recovery of the all-sky lensing power is as a result less biased compared to the alternative case. In short the binning of the mixing matrix should be preferably as fine as possible but the power spectrum needs to be rebinned for a better cosmological parameter estimation.

Overall, following the large-area bias correction, we find that the weak lensing convergence power spectrum can be both forward modelled and the full-sky convergence power reconstructed on scales greater than $l=2000$, for smaller-scale masking composed of both stars and a checkerboard (field-of-view) pattern. On scales less than $l=2000$ we see a slight residual excess of a few percent.

Propagating our lensing power spectra into  the error and bias on the cosmological parameters, $\sigma_8$ and $\Omega_m$, using the Fisher Matrix formalism for both Gaussian and lognormal fields, we find an unbiased measurement compared to the expected errors for a simulated survey of 100 deg$^2$. To estimate the Fisher matrix, the covariance matrix of the shear power is needed, and so in \App\ref{appLognormal} we provide a novel algorithm for calculating the moments of a lognormal field which we used to estimate the covariance of the power spectra for the simulated lognormal shear fields.

From our analysis we typically find a bias of around $1\%$ on parameters, while the error for 100 deg$^2$ is $3\%$ (with galaxy mean number density of 30/arcmin$^2$). Our results imply that this approach will remain unbiased for surveys up to 1200 deg$^2$. However, further study will be needed to improve this for larger surveys where the curvature of the sky will begin to be important.

In summary, we find that we can apply a flat-sky PCl analysis for a masked finite survey with stellar and checkerboard masks and that unbiased forward modelling and all-sky recovered convergence power can be recovered on angular scales above $l=2000$, with a few percent residual bias at lower wavenumbers. The flat-sky PCl method requires calibration to simulations to correct for an overall constant bias due to the survey geometry, but requires no other calibration. Both forward modelled and all-sky recovered power propagate into a small, percentage bias in measured cosmological parameter, which remain below the statistical accuracy for surveys of less than 1200 deg$^2$. Given similar results for forward modelling and all-sky recovery, there may be a slight preference for an all-sky recovery, since the forward modelling requires a convolution of the theory power spectrum at each point in parameter space.

Finally, we conclude that a flay-sky PCl method is suitable for the current generation of Weak Lensing surveys but, would be unsuited to surveys with high galaxy number density which are larger than 1200 deg$^2$, such as 15,000 deg$^2$ Euclid. The main bias seems to come from the large-area treatment of the survey geometry, where curved-sky effects will also become important. This could be alleviated with an all-sky spherical harmonic treatment. Given this can be slower, as FFT methods can only be used in the azimuthal directions, it may be that a hybrid method using flat-sky PCls on small-scales could be optimal.

\section*{Acknowledgements}
We thank Alan Heavens and Catherine Heymans for many great discussions, 
Lee Whittaker for discussions about curved sky and flat sky analysis and 
an anonymous referee for their constructive criticism. 
MA acknowledges support from the European Research Council, Scottish Universities Physics Alliance
and Euclid. ANT thanks the Royal Society for a Wolfson Research Merit Award.
BJ acknowledges support by an STFC Ernest Rutherford Fellowship, 
grant reference ST/J004421/1.
TDK is supported by a Royal Society University Research Fellowship.



\bibliographystyle{mnras}
\bibliography{biblio} 

\onecolumn
\appendix

\section{Mixing matrix}
\label{appMixing}

In \sect\ref{sec:Mask} we skipped some of the steps in calculating the mixing matrix. 
Here we show the details of the formalism \cite[based on][]{Yasin}.
The pseudo power spectrum can be written for EE, EB and BB correlations of the kappa map. 
In \sect\ref{sec:Mask} we started from \Eqt\eqref{eq:Cl1}, which shows the estimator used for
the power spectrum given the convergence on a finite patch of sky,
and derived \Eqt\eqref{eq:PCL1} which connects PCls to the underlying convergence maps.
Then we applied the ensemble averages to $\kappa_{\rm E,B}$ and used \Eqt\eqref{eq:Cl1}
to find a relation between the PCls and Cls. 
Here we write \Eqt\eqref{eq:Celldag1} for all combinations of the convergence maps,
\begin{align}
\label{eq:PCLAll2}
\left\langle \tC^{\rm EE}(\ell)\right\rangle =
 \frac{1}{A}\int\!\! \frac{\d\varphi_{\ell}}{2\pi}\!\!
 \int\!\!\frac{\d^2\ell'}{(2\pi)^2} |W(\lb-\lb')|^2
 \big\{C^{\rm EE}(\ell')\cos^2 2\varphi_{\ell \ell'}
 +[C^{\rm EB}(\ell')+C^{\rm BE}(\ell')]\sin 2\varphi_{\ell \ell'}\cos 2\varphi_{\ell \ell'}
 +C^{\rm BB}(\ell')\sin^2 2\varphi_{\ell \ell'}\big\}\;,\\ \nonumber
\left\langle \tC^{\rm EB}(\ell)\right\rangle =
 \frac{1}{A}\int\!\! \frac{\d\varphi_{\ell}}{2\pi}\!\!
 \int\!\!\frac{\d^2\ell'}{(2\pi)^2} |W(\lb-\lb')|^2
 \big\{C^{\rm EB}(\ell')\cos^2 2\varphi_{\ell \ell'}
 +[C^{\rm BB}(\ell')-C^{\rm EE}(\ell')]\sin 2\varphi_{\ell \ell'}\cos 2\varphi_{\ell \ell'}
 -C^{\rm BE}(\ell')\sin^2 2\varphi_{\ell \ell'}\big\}\;,\\ \nonumber
\left\langle \tC^{\rm BB}(\ell)\right\rangle =
 \frac{1}{A}\int\!\! \frac{\d\varphi_{\ell}}{2\pi}\!\!
 \int\!\!\frac{\d^2\ell'}{(2\pi)^2} |W(\lb-\lb')|^2
 \big\{C^{\rm BB}(\ell')\cos^2 2\varphi_{\ell \ell'}
 -[C^{\rm BE}(\ell')+C^{\rm EB}(\ell')]\sin 2\varphi_{\ell \ell'}\cos 2\varphi_{\ell \ell'}
 +C^{\rm EE}(\ell')\sin^2 2\varphi_{\ell \ell'}\big\}\;,
\end{align}
where $\varphi_{\ell\ell'}=\varphi_\ell-\varphi_{\ell'}$. 
Note that we have dropped $S(\ell)$ here as it is not used in the current work.
Since $C(\ell)$ do not have angular dependencies unlike $W(\lb-\lb')$ and the trigonometric functions,
we can take the integrals over $\varphi$ and $\varphi'$ separately. 
To do so we first write,
\begin{equation}
\label{eq:Wll}
 |W(\lb-\lb')|^2=2\pirm\int_0^\infty \d L L \int_0^{2\pirm}\frac{\d \varphi_{L}}{2\pirm} 
 |W(\Lb)|^2 \delta_{\rm D}\big(\Lb-(\lb-\lb')\big)\;,
\end{equation}
where $\delta_{\rm D}$ is the Dirac delta function, which we exchange with its integral form,
\begin{equation}
 \delta_{\rm D}\big(\Lb-(\lb-\lb')\big)=
 \int \frac{\d^2 \theta}{(2\rm\pi)^2}{\rm e}^{-\i\tb.\left(\Lb-(\lb-\lb')\right)}\;,
\end{equation}
and define,
\begin{equation}
W_{\upgamma\upgamma}(L)\equiv\int_0^{2\rm\pi}\frac{\d\varphi_{L}}{2\pi}|W(\boldsymbol L)|^2\;.
\end{equation}
Substituting for $|W(\lb-\lb')|^2$ in \Eqt\eqref{eq:PCLAll2} leads to,
\begin{align}
 \label{eq:PCLAll3}
\left\langle \tC^{\rm EE}(\ell)\right\rangle &= \frac{1}{A}\int_0^\infty  \frac{\d \ell'\ell'}{(2 \rm \pi)^2}
  \int_{|\ell-\ell'|}^{|\ell+\ell'|}  \d L L W_{\upgamma\upgamma}(L)
  \left(\int \d^2 \theta {\rm e}^{-\i\tb.\Lb}\right)\\ \nonumber
  &\times\int_0^{2\pirm} \frac{\d \varphi_\ell}{2 \rm \pi}
  \int_0^{2\pirm} \frac{\d \varphi_{\ell'}}{2 \rm \pi}
  {\rm e}^{\i\tb.\lb}{\rm e}^{-\i\tb.\lb'}
  \big\{C^{\rm EE}(\ell')\cos^2 2\varphi_{\ell \ell'}
  +[C^{\rm EB}(\ell')+C^{\rm BE}(\ell')]\sin 2\varphi_{\ell \ell'}\cos 2\varphi_{\ell \ell'}
  +C^{\rm BB}(\ell')\sin^2 2\varphi_{\ell \ell'}\big\}\;,\\ \nonumber
\left\langle \tC^{\rm EB}(\ell)\right\rangle &= \frac{1}{A}\int_0^\infty \frac{\d \ell'\ell'}{(2 \rm \pi)^2}
  \int_{|\ell-\ell'|}^{|\ell+\ell'|} \d L L W_{\upgamma\upgamma}(L)
  \left(\int \d^2 \theta {\rm e}^{-\i\tb.\Lb}\right)\\ \nonumber
  &\times\int_0^{2\pirm} \frac{\d \varphi_\ell}{2 \rm \pi}
  \int_0^{2\pirm} \frac{\d \varphi_{\ell'}}{2 \rm \pi}
  {\rm e}^{\i\tb.\lb}{\rm e}^{-\i\tb.\lb'}
  \big\{C^{\rm EB}(\ell')\cos^2 2\varphi_{\ell \ell'}
 +[C^{\rm BB}(\ell')-C^{\rm EE}(\ell')]\sin 2\varphi_{\ell \ell'}\cos 2\varphi_{\ell \ell'}
 -C^{\rm BE}(\ell')\sin^2 2\varphi_{\ell \ell'}\big\}\;,\\ \nonumber
\left\langle \tC^{\rm BB}(\ell)\right\rangle &= \frac{1}{A}\int_0^\infty \frac{\d \ell'\ell'}{(2 \rm \pi)^2}
  \int_{|\ell-\ell'|}^{|\ell+\ell'|} \d L L W_{\upgamma\upgamma}(L)
  \left(\int \d^2 \theta {\rm e}^{-\i\tb.\Lb}\right)\\ \nonumber
  &\times\int_0^{2\pirm} \frac{\d \varphi_\ell}{2 \rm \pi}
  \int_0^{2\pirm} \frac{\d \varphi_{\ell'}}{2 \rm \pi}
  {\rm e}^{\i\tb.\lb}{\rm e}^{-\i\tb.\lb'}
  \big\{C^{\rm BB}(\ell')\cos^2 2\varphi_{\ell \ell'}
 -[C^{\rm BE}(\ell')+C^{\rm EB}(\ell')]\sin 2\varphi_{\ell \ell'}\cos 2\varphi_{\ell \ell'}
 +C^{\rm EE}(\ell')\sin^2 2\varphi_{\ell \ell'}\big\}\;.
\end{align}
The angular dependency of the integral in parentheses can be taken separately to yield,
\begin{equation}
\int \d^2 \theta {\rm e}^{-\i\tb.\Lb}=2{\rm\pi}\int_0^\infty \d\theta \theta J_{0}(L\theta)\;,
\end{equation}
where $J_{0}$ is the zeroth order Bessel function of the first kind. 
We can now take the integrals over $\varphi_\ell$ and $\varphi_\ell'$,
using the following relations,
\begin{equation}
J_n(x)=\frac{1}{2\pirm \i^n}\int_0^{2\pirm}\d\varphi\; {\rm e}^{\i x \cos\varphi}{\rm e}^{\i n \varphi}\;, 
~~~~~~ J_{-n}(x)=(-1)^n J_n(x)\;,
\end{equation}
and
\begin{equation}
\cos\varphi=\frac{{\rm e}^{\i\varphi}+{\rm e}^{-\i\varphi}}{2}\;,
~~~~~~~ \sin\varphi=\frac{{\rm e}^{\i\varphi}-{\rm e}^{-\i\varphi}}{2i}\;,
\end{equation}
which result in these equations,
\begin{align}
&\int_0^{2\pirm} \frac{\d \varphi_\ell}{2 \rm \pi}\int_0^{2\pirm} \frac{\d \varphi_{\ell'}}{2 \rm \pi}
  {\rm e}^{\i\theta\ell\cos\varphi_\ell}{\rm e}^{-\i\theta\ell'\cos\varphi_{\ell'}}\cos^2 2\varphi_{\ell\ell'}=
  \frac{1}{2}\left[J_0(\ell\theta)J_0(\ell'\theta)+J_4(\ell\theta)J_4(\ell'\theta)\right]\;,\\
&\int_0^{2\pirm} \frac{\d \varphi_\ell}{2 \rm \pi}\int_0^{2\pirm} \frac{\d \varphi_{\ell'}}{2 \rm \pi}
  {\rm e}^{\i\theta\ell\cos\varphi_\ell}{\rm e}^{-\i\theta\ell'\cos\varphi_{\ell'}}\sin^2 2\varphi_{\ell\ell'}=
  \frac{1}{2}\left[J_0(\ell\theta)J_0(\ell'\theta)-J_4(\ell\theta)J_4(\ell'\theta)\right]\;,\\
&\int_0^{2\pirm} \frac{\d \varphi_\ell}{2 \rm \pi}\int_0^{2\pirm} \frac{\d \varphi_{\ell'}}{2 \rm \pi}
  {\rm e}^{\i\theta\ell\cos\varphi_\ell}{\rm e}^{-\i\theta\ell'\cos\varphi_{\ell'}}
  \cos 2\varphi_{\ell\ell'}\sin 2\varphi_{\ell\ell'}=0\;,
\end{align}
where $J_4$ is the fourth order Bessel function of the first kind. 
After these simplifications we are left with combinations of three Bessel functions which are the only functions
that depend on $\theta$. We then find analytic solutions for the integrals over $\theta$ using, 
\begin{equation}
\int_0^\infty \d \theta\;\theta J_0(L\theta)J_n(\ell\theta)J_n(\ell'\theta)=
\frac{\cos n\eta}{{\rm\pi}\ell\ell'\sin\eta}\;,
\end{equation}
where $\eta$ is the angle between $\ell$ and $\ell'$ (see Gradshteyn and Ryzhik 1994). 
Note that by definition $\Lb=\lb-\lb'$ (see \Eqt\ref{eq:Wll}), 
which means they form a triangle of area $\frac{1}{2}\ell\ell'\sin\eta$.
Substituting for the $\theta$, $\varphi_\ell$ and $\varphi_{\ell'}$  integrals in \Eqt\eqref{eq:PCLAll3} we find
\begin{align}
\label{eq:PCLAllEnd}
\left\langle \tC^{\rm EE}(\ell)\right\rangle &= \frac{1}{A}\int_0^\infty \frac{\d \ell'\ell'}{(2 \rm \pi)^2}
  \int_0^\pi \d\eta \; W_{\upgamma\upgamma}(L) 
 \big\{(1+\cos4\eta)C^{\rm EE}(\ell')+(1-\cos4\eta)C^{\rm BB}(\ell')\big\}\;,\\ \nonumber
\left\langle \tC^{\rm EB}(\ell)\right\rangle &= \frac{1}{A}\int_0^\infty  \frac{\d \ell'\ell'}{(2 \rm \pi)^2}
  \int_0^\pi  \d\eta\; W_{\upgamma\upgamma}(L) (2\cos4\eta) C^{\rm EB}(\ell')\;,\\ \nonumber
\left\langle \tC^{\rm BB}(\ell)\right\rangle &= \frac{1}{A}\int_0^\infty  \frac{\d \ell'\ell'}{(2 \rm \pi)^2}
 \int_0^\pi \d\eta\;  W_{\upgamma\upgamma}(L)
 \big\{(1-\cos4\eta)C^{\rm EE}(\ell')+(1+\cos4\eta)C^{\rm BB}(\ell')\big\}\;,\\ \nonumber
\end{align}
where we used $L^2=\ell^2+\ell'^2-2\ell\ell'\cos\eta$ to replace $\d L L/(\ell\ell'\sin\eta)$ with $\d\eta$.
\Eqt\eqref{eq:PCLAllEnd} shows that the EB power spectrum does not mix with EE and BB. 
Consequently, we ignored this term in
\sect\ref{sec:Mask}. Note that the above calculations are accurate for 
an idealistic case where all the angles are available. 
The mixing matrix is then formed directly from the above equations.

\section{Power spectrum plots}
\label{appClPlots}

The $C(\ell)$ and $\tC(\ell)$ plots  for the composite mask were shown in \fig\ref{figPCL}. 
Similar plots for all the
masks are shown in this appendix, including a control case without a mask. 
All the cases are zero-padded before the measurements, even the control case. 
In addition, we show plots of estimated to theory ratios for both the 
recovered $C(\ell)$ and $\tC(\ell)$, for all the masks.
These plots were shown for the composite mask in \fig\ref{figCLRatio}. 

In total we have 4 mask types: ``No Mask'', ``Star'', ``Checkerboard'' and ``Composite'', as well as 4 types of 
apodisation: `` No Ap'', ``Ap1'', ``Ap2'' and ``Ap3''. 
``No Ap'' means no apodisation was applied and the rest of the apodisation
options are explained in \sect\ref{sec:Apodise}. The masks are discussed in \sect\ref{sec:Mask}. 

\begin{figure}
  \begin{center}
    \begin{tabular}{c}
      \resizebox{170mm}{!}{\includegraphics{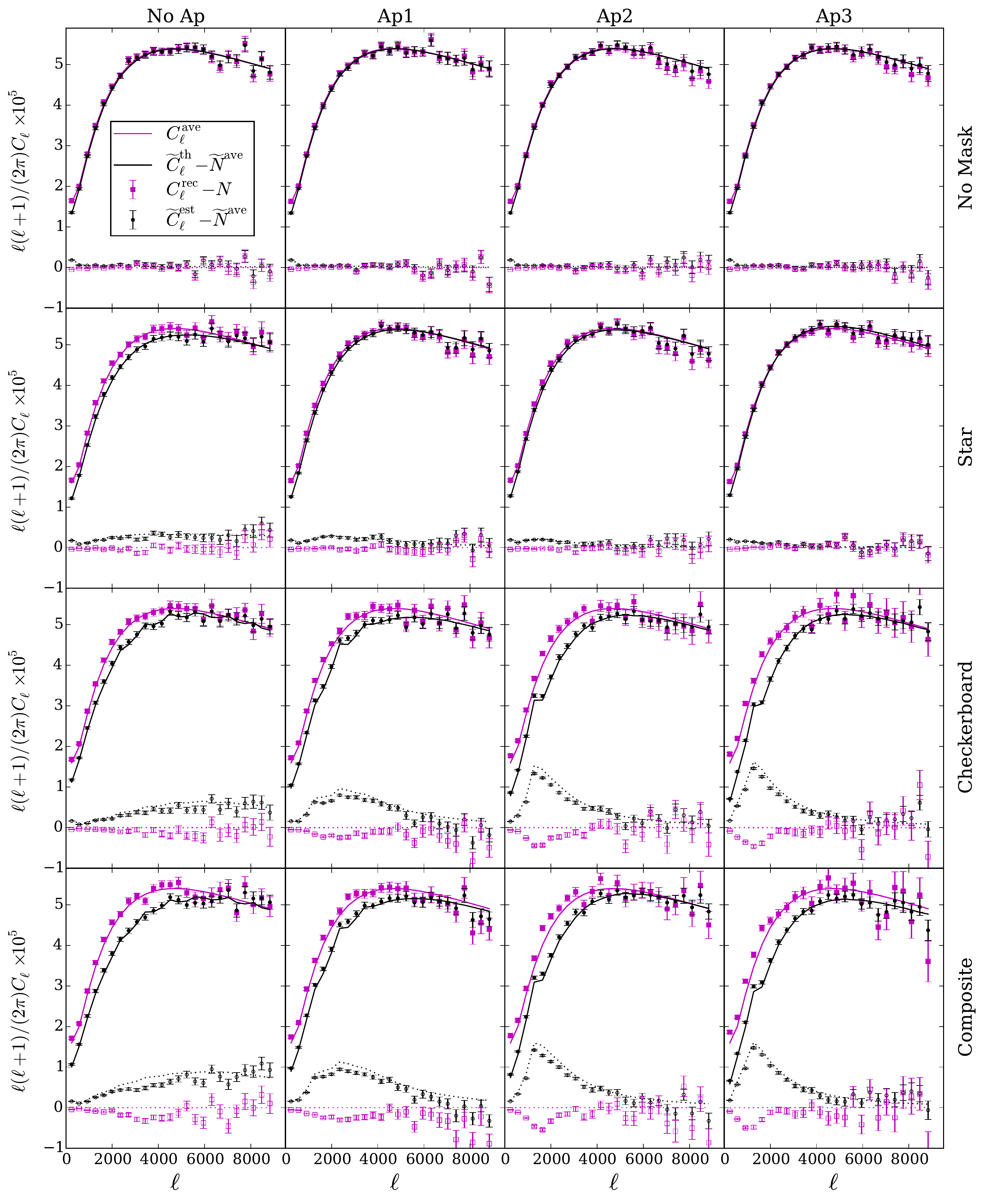}}  
    \end{tabular}
     \caption{\small{$C(\ell)$ and $\tC(\ell)$ plots for all masks and apodisation configurations. 
     The magenta solid curve shows the input angle-averaged power spectrum. 
     The magenta solid and open squares show the recovered $C(\ell)$ from method one 
     (see \Eqt\ref{eq:ClRecI}) for E/B-modes respectively.
     The solid and dotted black curves show the theory values of the $\tC(\ell)$ for E/B-modes, 
     while the black solid and open circles show their estimated value from the simulated fields. 
     The noise contribution is subtracted here. The error-bars show the variance of the mean of the 100 fields. 
     The columns show the apodisation used, whereas the rows show the mask type used for each plot. 
     `` No Ap'' and ``No Mask'' mean no apodisation and no mask was used. The first row is used as the control case.} }
    \label{figPCLAll}
  \end{center}
\end{figure}

\fig\ref{figPCLAll} shows the estimated and theory values of $C(\ell)$ 
and $\tC(\ell)$ for all the mask configurations with a wide binning ($n_\ell=20$, $\Delta\ell\approx 360$).
In \fig\ref{figPCLAll} the magenta curves and squares show the theory and recovered $C(\ell)$, 
while the black curves and circles show the theory and estimated $\tC(\ell)$ (explained in more detail in the caption).
The control case with no masks shows very little difference between the $C(\ell)$ and $\tC(\ell)$ for most scales. 
The largest difference is at very large and very small scales. 
At very large scales the difference is due to the zero-padding, which effectively acts as a large scale mask.
The small scale differences appear at the scales where noise is dominant ($\ell \gtrsim 5000$). 
Additionally, we see more fluctuations at these scales for $\tC(\ell)^{\rm est}$ and  $C(\ell)^{\rm rec}$ as expected.
The apodisation has very little effect on the no mask case.

The second row of \fig\ref{figPCLAll} shows the results for the star mask, i.e. small circular masks. 
Looking at the left most plot in this row, the ``No Ap'' case, 
we see that the overall effect of the star mask is to lower the amplitude of the small to midrange E-mode $\tC(\ell)$, 
by shifting the power to B-modes.
Apodisation moves the $\ell$-modes at which the leakage takes place,
which is directly related to the size of the smoothing kernel 
(a larger kernel stops leakage for a larger range of $\ell$).  
As a result the Star and Ap3 plot resembles the control plots. 

The third row of \fig\ref{figPCLAll} shows the results for the checkerboard mask, i.e. large CCD patterns.
Similar to the star mask, the overall effect is a leakage of E-modes into B-modes which decreases the amplitude of 
the E-mode $\tC(\ell)$. However, the regular patterns in this mask produce structures in the  $\tC(\ell)$.
Apodisation makes the resulting $\tC(\ell)$ smoother at large $\ell$ and pushes the structures to smaller $\ell$-modes. 
The mask modelling fails to capture the structures accurately 
in the presence of apodisation, as can be seen in the plots.

Finally the last row of \fig\ref{figPCLAll} shows the results for the composite mask. As both components of this mask 
reduce the amplitude of the E-modes by moving the power to B-modes, the effect is more pronounced for the composite. 
The structures of the checkerboard mask can also be seen here. 
The B-mode modelling is poor at certain scales for this mask.
These scales are pushed to lower $\ell$-modes with apodisation.

\begin{figure}
  \begin{center}
    \begin{tabular}{c}
      \resizebox{170mm}{!}{\includegraphics{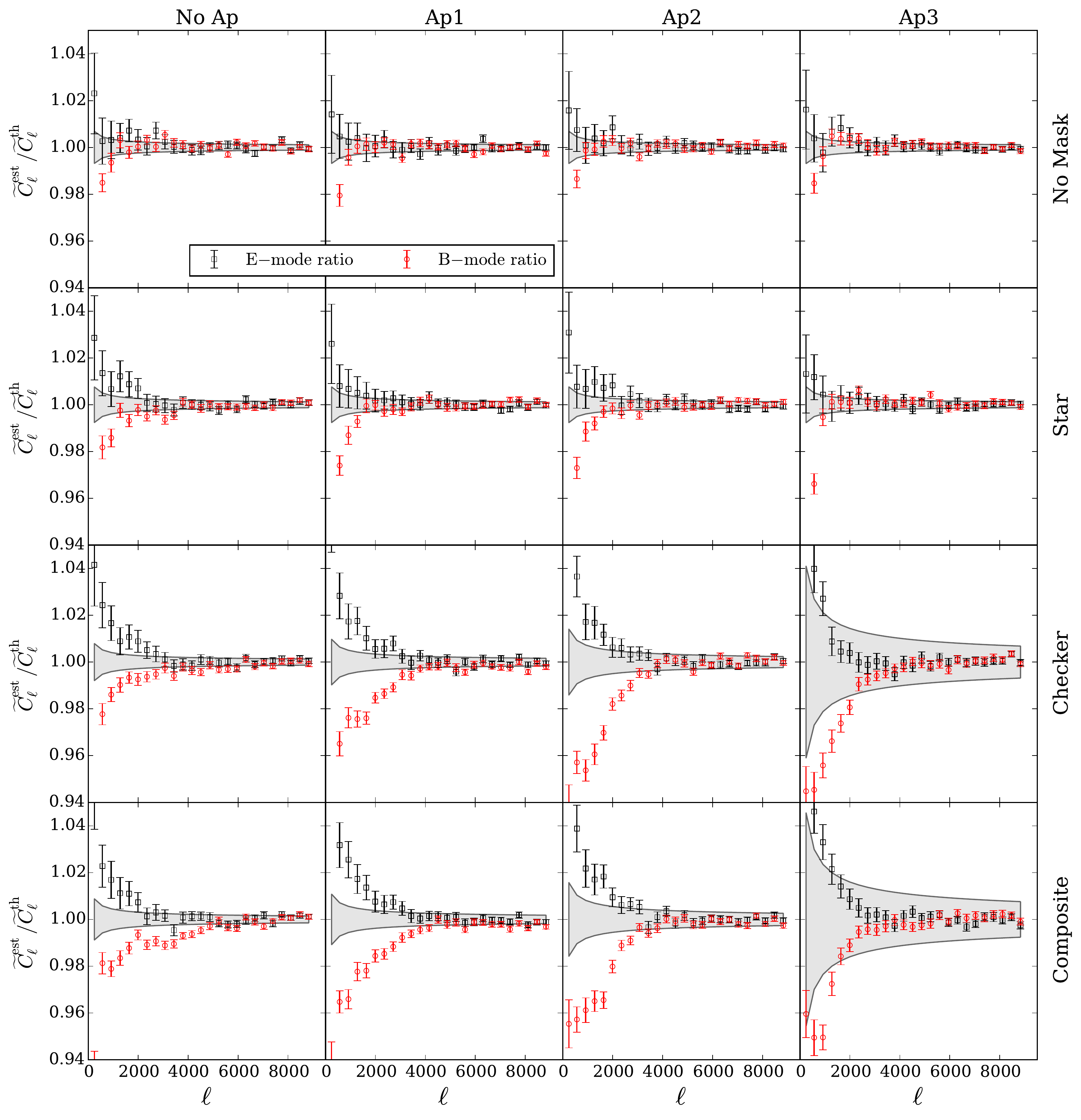}}  
    \end{tabular}
     \caption{\small{$\tC(\ell)$ ratio plots for all masks and apodisation configurations. 
     The black squares show the ratio for the E-modes, while the red circles belong to B-modes.
     The grey shaded area shows the expected cosmic variance centred at one. 
     We expect to find a good agreement between the estimated and the theory values within the cosmic variance band 
     if the mask modelling is accurate. 
     The error-bars show the variance of the mean and are estimated from the simulations.} }
    \label{figPCLRatioAll}
  \end{center}
\end{figure}

In order to objectively investigate the mask modelling and its limitations 
we need to look at ratio plots rather than \fig\ref{figPCLAll}, 
since the error-bars are very small which makes any judgement from 
\fig\ref{figPCLAll} difficult.
Hence \fig\ref{figPCLRatioAll} shows the ratio of the estimated $\tC(\ell)$ 
to its theory value for all the mask and apodisation configurations.
The y-range here is different from \fig\ref{figCLRatio}, for better inspection. 
The noise contribution is not subtracted from $\tC(\ell)$.
The plots are shown for the lognormal fields and the largest binning ($\ell \gtrsim 5000$). 
The red circles show the ratio of the B-modes and the black squares the E-modes. 
The grey shaded area shows the expected cosmic variance contribution
for each case (see \Eqt\ref{eq:CosmicVar2}). 
The control case with no masking shows discrepancies at small $\ell$ between the estimated 
and the theory $\tC(\ell)$ specially for the B-modes. 
However, the rest is within the cosmic variance band. 
The star mask cases also show a similar behaviour, whereas the checkerboard 
mask pushes the small $\ell$ discrepancies to larger values. 
The apodised checkerboard mask covers a much larger area of 
the field compared to the non-apodised version, specially for Ap3. 
Consequently, the cosmic variance increases rapidly for this mask with larger smoothing kernels. 
The discrepancies seen for the star and checkerboard
masks add up for the composite mask. 

\begin{figure}
  \begin{center}
    \begin{tabular}{c}
      \resizebox{170mm}{!}{\includegraphics{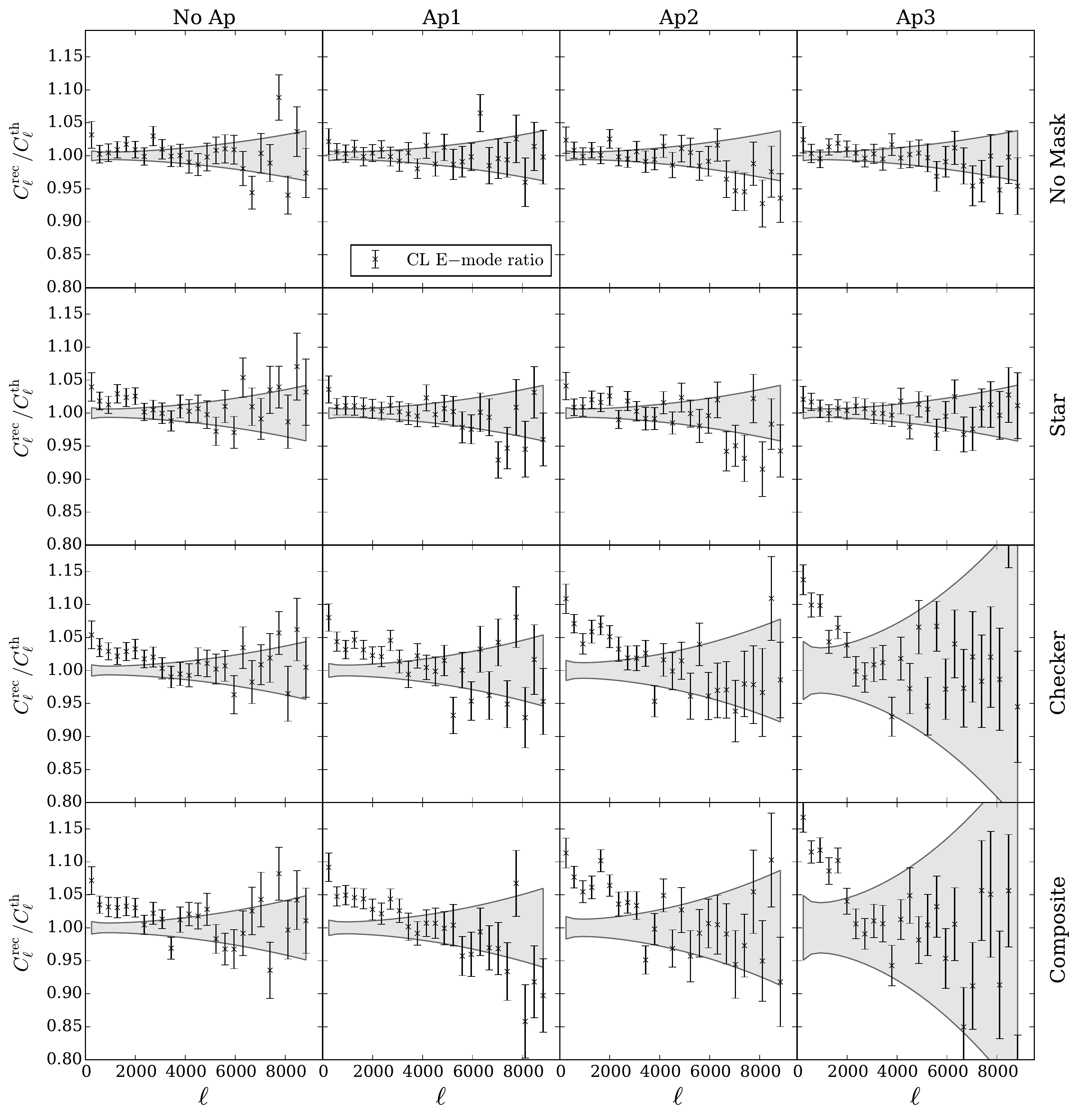}}  
    \end{tabular}
     \caption{\small{$C(\ell)$ ratio plots for all masks and apodisation configurations. 
     The symbols show the ratios for the recovered E-mode $C(\ell)$ to the input angle averaged $C(\ell)$.
     The grey area shows the cosmic variance contribution, which widens with an increased masked area.
     The error on the mean is estimated from the field-to-field variance of 100 lognormal simulations.} }
    \label{figCLRatioAll}
  \end{center}
\end{figure}

\fig\ref{figCLRatioAll} shows the ratio of the $C(\ell)^{\rm rec}$ 
in \Eqt\eqref{eq:ClRecI} (method I) to the input $C(\ell)^{\rm ave}$. 
The noise contribution has been subtracted from the recovered $C(\ell)$.
The grey shaded area shows the cosmic variance centred on one. 
The No Mask row used as the control case, shows a good agreement between 
the theory and recovered values. The Star cases show a similar behaviour except 
for a slightly overestimated recovery of the lowest $\ell$-mode.
The checkerboard cases show disagreements up to $\ell\approx2000$, which can also be seen for the composite mask.

\section{Lognormal moments}
\label{appLognormal}
We can find the moments of a lognormal field using its relation to a Gaussian field. 
These moments will be used to calculate the covariance matrix of the power spectrum of the 
lognormal realisation, which will then be used to estimate the Fisher matrices.
In this Appendix we will show how all the moments of a lognormal field can be written  in terms of 
its power spectrum, by taking the following steps.

A lognormal field, $\dlg$, is defined with respect to a Gaussian field, $\delta$, with zero mean, as
\begin{equation}
\label{eq:LogNormal}
 \dlg({\bf x})\equiv  e^{\delta({\bf x})-\sigma^2/2}-1\;,
\end{equation}
where $\sigma^2$ is the variance of the Gaussian field. 
This implies that the variance of the lognormal field is $\exp\sigma^2$.
To find the moments of this lognormal field, we take the following steps. 
In Fourier space the Nth lognormal moment for a non-zero $\kb_i$ can be written as
\begin{equation}
\label{eq:dlgh}
 \Big\langle\prod_i^N\dlgh(\kb_i)\Big\rangle 
 =\Big\langle \prod_i^N[(2\pi)^n\delta_{\rm D}(\kb_i)+\dlgh(\kb_i)]\Big\rangle\;,
\end{equation}
where $\delta_{\rm D}(\kb_i)$ is the Dirac delta function.
Note that $\kb_i$ has $n$ dimensions and
$\delta(\kb_i)$ is a one dimensional quantity on a multidimensional grid. 
Next we write the lognormal moments with respect to their real space counterparts,
\begin{equation}
 \Big\langle \prod_i^N[(2\pi)^n\delta_{\rm D}(\kb_i)+\dlgh(\kb_i)]\Big\rangle= 
 \Big\langle \prod_i^N\int\d \xb_i e^{-\i\kb_i.\xb_i}[1+\dlg_i]\Big\rangle\;,
\end{equation}
where $\dlg_i\equiv\dlg(\xb_i)$.
We can take the ensemble average inside the integral and rewrite the above equation as,
\begin{equation}
\label{eq:AveFourier}
 \Big\langle \prod_i^N\dlgh(\kb_i)\Big\rangle= 
 \prod_i^N\left[\int\d \xb_i e^{-\i\kb_i.\xb_i}\right]
 \Big\langle \prod_i^N[1+\dlg_i]\Big\rangle  \;.
\end{equation}
The ensemble average, $\langle \prod_i^N[1+\dlg_i]\rangle$, 
can be expressed in terms of the two point correlation functions,
by writing the lognormal fields in terms of their Gaussian generators from \Eqt\eqref{eq:LogNormal}, 
\begin{align}
\label{eq:Meanlg}
 \Big\langle \prod_i^N(1+\dlg_i)\Big\rangle =  
 \int_{-\infty}^{+\infty}\d\deltab\: p(\deltab) \;
   \prod_i^N \left[e^{\delta_i}\;e^{-\sigma_i^2/2}\right] \;,
\end{align}
where $\deltab\equiv(\delta_1,\delta_2,...,\delta_N)$ and
\begin{equation}
 p(\deltab)=\frac{e^{-\deltab\; \CM^{-1}\deltab^t/2}}{\sqrt{(2\pi)^N\det \CM}}\;,
\end{equation}
is the multivariate Gaussian distributed probability of $\deltab$ with $\CM$ as the covariance. 

Since $\CM$ is a covariance, i.e. symmetric and positive definite, 
we can write it in terms of its eigenvalues and eigenvectors as 
\begin{equation}
 \CM=\OM \DM \OM^{\rm t}\;,
\end{equation}
where $\OM$ is the orthogonal matrix made out of the eigenvectors 
of $\CM$ and $\DM$ is a diagonal matrix of the eigenvalues of $\CM$.
As a result 
\begin{equation}
 \CM^{-1}=\OM \DM^{-1} \OM^{\rm t}\;.
\end{equation}
Substituting for $\CM^{-1}$ in \Eqt\eqref{eq:Meanlg} and defining $ X\equiv\deltab\OM$, yields,
\begin{align}
\label{eq:Meanlg2}
 \Big\langle \prod_i^N(1+\dlg_i)\Big\rangle =  
    \frac{\prod_i^N e^{-\sigma_i^2/2}}{\sqrt{(2\pi)^N\det \CM}}
    \int_{-\infty}^{+\infty}\d X\: e^{-X \DM^{-1}X^t/2}\;
    \prod_i^N \left[e^{X_j O^{\rm t}_{j i}}\;\right] \;,
\end{align}
where $|\det\OM|=1$ was used to simplify the result.
Rewriting \Eqt\eqref{eq:Meanlg2} in terms of its components results in,
\begin{align}
\label{eq:Meanlg3}
 \Big\langle \prod_i^N (1+\dlg_i)\Big\rangle =  
    \frac{\prod_i^N e^{-\sigma_i^2/2}}{\sqrt{(2\pi)^N\det \CM}}
    \prod_j^N     \left[\int_{-\infty}^{+\infty}\d X_{j}\: e^{-X_j^2 D_{jj}^{-1}/2+X_j\sum_i O_{ij}}\right]\;.
\end{align}
With the aid of another variable change, $Y_i=X_i\sqrt{D_i^{-1}}$, 
and completing the square we solve this integral and find the desired relationship,
\begin{align}
\label{eq:Meanlgfinal}
 \Big\langle \prod_i^N(1+\dlg_i)\Big\rangle = 
 \exp\left( \frac{\sum_{i j} C_{ij}-\sigma^2_i}{2}\right)=
 \exp\left( \sum_{i<j} C_{ij}\right).
\end{align}
Inserting for $\langle \prod_i^N\dlgh(k_i)\rangle$ from the above equation 
into \Eqt\eqref{eq:AveFourier} results in,
\begin{equation}
\label{eq:AveFourier2}
 \langle \prod_i^N\dlgh(k_i)\rangle= \prod_i^N\int\d x_i e^{-\i k_i.x_i}\prod_{i<j}^Ne^{C_{ij}}  \;.
\end{equation}
The covariance of the Gaussian and lognormal fields are related via,
\begin{equation}
e^{C_{ij}}=1+\langle \delta_i^{\rm ln}\delta_j^{\rm ln}\rangle=1+\xi_{ij}^{\rm ln}\;,
\end{equation}
where $\xi^{\rm ln}_{ij}$ is the correlation between $\delta^{\rm ln}_i$ and $\delta^{\rm ln}_j$. 
Consequently, we can write the lognormal moments in \Eqt\eqref{eq:AveFourier2}
in terms of their two point correlation functions
or alternatively their power spectra,
\begin{align}
\label{eq:AveFourier3}
 \langle \prod_i^N\dlgh(k_i)\rangle= \frac{1}{(2\pi)^{M n}}\prod_i^N\int\d x_i e^{-\i k_i.x_i}
\prod_{j>i;m}^{N;M} \int \d l_m e^{\i l_m.(x_i-x_j)} [(2\pi)^n\delta_{\rm D}+P](l_m)  \;,
\end{align}
where $P(l)$ is the power spectrum of the lognormal field, $n$ is the dimension of the field 
(for gravitational lensing $n=2$), the subscript $m$ belongs 
to each pair of $\delta_i^{\rm ln}$ and $\delta_j^{\rm ln}$ 
which make $\langle \delta_i^{\rm ln}\delta_j^{\rm ln}\rangle=\xi_{ij}^{\rm ln}$ and $M=N(N-1)/2$.
The above integrals can be simplified by integrating with respect to $x_i$, 
since $[(2\pi)^n\delta_{\rm D}+P](l_m)$ have no dependency on $x_i$. 
The $x_i$ integrals will result in $N$ delta functions of dimension $n$ which depend on 
$k_i$ and $l_m$. 
There are $M$, $l_m$ integrals and $2^M$, $[(2\pi)^n\delta_{\rm D}(l_m)+P(l_m)]$ combinations. 
Writing the Delta functions found from the $x_i$ integrals in the following form,
\begin{align}
 \delta_i^{\bar{\sum} j-\bar{\sum} k}&\equiv\delta_i^{+j+j'+...-k-k'-...}
 \equiv\delta_{\rm D}(k_i+l_j+l_{j'}+...-l_k-l_{k'}-...)\;,
\end{align}
will simplify the notation. Note that $\bar{\sum}$ is not a real sum.
We find that the two sums over the positive and negative $l_m$ modes can be formulated as follows,
\begin{align}
 \bar{\sum} j ={\bar{\sum}_{r=1}^{i-1}}(r-1)N+i-r(r+1)/2 
 -\bar{\sum} k = -{\bar{\sum}_{r=i}^{N-1}}(i-1)N-i(i-1)/2+r-i+1\;,
\end{align}
for a given $N$ and $i$. We are now left with $M$ integrals with $2^M$ components for each,
\begin{align}
 \langle \prod_i^N\dlgh(k_i)\rangle=\frac{1}{(2\pi)^{n(M-N)}}\prod_{j>i;m}^{N;M} 
 \int \d l_m[(2\pi)^n\delta_{\rm D}+P](l_m)
\prod_i^N\delta_i^{+\bar{\sum}_{r=1}^{i-1}(r-1)N+i-r(r+1)/2
-{\bar{\sum}_{r=i}^{N-1}}(i-1)N-i(i-1)/2+r-i+1}\;.
\end{align}
The remaining $M$ integrals over $l_m$ can be simplified using the $N$ delta functions 
and the delta functions in the $2^M$ combinations of $\delta_{\rm D}(l_m)$ and $P(l_m)$. 
Some of these integral vanish after considering the delta functions. 
In any integral if we come about a $\delta_{\rm D}(k_i)$ then that term is 
equal to zero since we are not interested in $k_i=0$ terms and for the rest 
of the values the delta function vanishes.

We can immediately see that for the third moment $N=M=3$,
i.e. only one integral will remain after the simplifications 
and the rest of the term will either vanish or are products of power spectra and
Delta functions which depend on several $k_i$ modes. 
We have developed an algorithm which can simplify the moments for any given $N$.

The fourth moment of the lognormal fields are essential for calculating the
covariance of their power spectra. Therefore, here the results for the fourth order moment will be
explicitly shown. The fourth moment has many terms. These terms can be divided into four groups,
depending on the number of remaining integrals over the power spectra and an extra group which 
contains the Gaussian only contribution. Hence, in the following each group will be represented 
separately. 
The fourth lognormal moment in Fourier space can be written as,
\begin{align}
\label{eq:4thMoment}
\langle \dlgh(k_1)\dlgh(k_2)\dlgh(k_3)\dlgh(k_4)\rangle=
(2\pi)^n\dD(k_1+k_2+k_3+k_4)\{I+II+III+IV\}+ G\;,
\end{align}
where $G$ is the pure Gaussian term,
\begin{align}
G=(2\pi)^{2n}\big[&\dD(k_2+k_3) \dD(k_1+k_4)  P(k_1) P(k_2)\\ \nonumber
+&\dD(k_1+k_3) \dD(k_2+k_4)  P(k_1) P(k_2)\\ \nonumber
+&\dD(k_1+k_2) \dD(k_3+k_4)  P(k_1) P(k_3)\big]\;,   
\end{align}
and $I$, $II$, $III$ and $IV$ are the pure lognormal terms, shown bellow. 
The highest number of integrals remaining after the simplifications is three.
There is only a single term of this form,
\begin{align}
\label{eq:3Int}
I=\int\d l_4\d l_5\d l_6 P(l_4)P(l_5)P(l_6) P(l_4+l_5-k_2) P(l_6-l_4-k_3)
P(k_4+l_5+l_6)\;.
\end{align}
There are six terms with two integrals, which can be factorized as,
\begin{align}
II= &\int  \d l_5  \d l_6 P(l_5) P(l_6) P(k_4\!+\!l_5\!+\!l_6) 
\big[ P(l_5\!+\!l_6\!-\!k_2\!-\!k_3) P(l_5\!-\!k_2)
+P(l_5\!+\!l_6\!-\!k_2\!-\!k_3)  P(l_6\!-\!k_3)
+P(l_5\!-\!k_2) P(l_6\!-\!k_3)  \big]\nonumber \\ 
+&\int  \d l_4  \d l_6 P(l_4) P(l_6) P(l_6\!-\!l_4\!-\!k_3)P(k_4\!+\!l_6) 
\big[P(l_4\!-\!l_6\!+\!k_1\!+\!k_3) +  P(l_4\!-\!k_2)  \big]\\ \nonumber
+&\int  \d l_4  \d l_5 P(l_4) P(l_5) P(l_4\!+\!l_5\!-\!k_2) P(l_4+k_3) P(k_4\!+\!l_5)\;. 
\end{align}
The fifteen terms that have one remaining integral are,
\begin{align}
III&=\int\d l_6 P(l_6) \big[P(k_1) P(l_6\!-\!k_3) P(l_6\!-\!k_2\!-\!k_3) 
+P(k_1) P(k_4\!+\!l_6) P(l_6\!-\!k_1\!-\!k_3)  \\ \nonumber
&~~~~~~~~~~~~~~~~~+  P(k_1) P(k_4\!+\!l_6) P(l_6\!-\!k_3)  
+P(k_2) P(l_6\!-\!k_3) P(l_6\!-\!k_1\!-\!k_3)  \\ \nonumber
&~~~~~~~~~~~~~~~~~+  P(k_2) P(k_4\!+\!l_6) P(l_6\!-\!k_2\!-\!k_3) 
 +  P(k_2) P(k_4\!+\!l_6)  P(l_6\!-\!k_3) \\ \nonumber
&~~~~~~~~~~~~~~~~~+  P(k_4\!+\!l_6) P(l_6\!-\!k_2\!-\!k_3)  P(l_6\!-\!k_3) 
+P(k_4\!+\!l_6) P(l_6\!-\!k_1\!-\!k_3) P(l_6\!-\!k_3) \big]\\ \nonumber
&+ \int  \d l_5 P(l_5) \big[P(k_4\!+\!l_5) P(l_5\!-\!k_2) P(l_5\!-\!k_2\!-\!k_3) 
+  P(k_3) P(l_5\!-\!k_2) P(l_5\!-\!k_1\!-\!k_2)  \\ \nonumber
&~~~~~~~~~~~~~~~~~+  P(k_3) P(k_4\!+\!l_5) P(l_5\!-\!k_2\!-\!k_3)  
+  P(k_3) P(l_5\!-\!k_2)  P(k_4\!+\!l_5) \big]\\ \nonumber
&+ \int  \d l_4 P(l_4)P(k_4)\big[P(l_4\!-\!k_2) P(l_4\!-\!k_1\!-\!k_2)  
+  P(l_4\!+\!k_3) P(l_4\!+\!k_1\!+\!k_3)+  P(l_4\!-\!k_2) P(l_4\!+\!k_3)  \big]\;. \nonumber
\end{align}
And finally there are 16 terms which do not have any remaining integrals and only depend on
the power spectra of the lognormal modes,
\begin{align}
IV&=P(k_1)P(k_2)P(k_3)+P(k_1)P(k_2)P(k_4)
+ P(k_1)P(k_3)P(k_4)+P(k_2)P(k_3)P(k_4)\\  \nonumber
&+ [P(k_1)P(k_2)+P(k_3)P(k_4)] [P(k_1+k_3)+P(k_2+k_3)]\\  \nonumber 
&+ [P(k_1)P(k_3)+P(k_2)P(k_4)][P(k_1+k_2)+ P(k_2+k_3)]\\  \nonumber
&+ [P(k_1)P(k_4)+P(k_2)P(k_3)][P(k_1+k_2)+ P(k_1+k_3)]\;.\\  \nonumber
\end{align}
$IV$ has the highest contribution out of all of the pure lognormal terms as was shown by 
\cite{HHS11} for the covariance of the two point correlation functions.
Ergo, to find the covariance of the power spectra for a lognormal field we neglect 
$I$, $II$ and $III$.

\bsp	
\label{lastpage}

\end{document}